\documentclass[iop,apj,twocolappendix, numberedappendix]{emulateapj}

\usepackage{amsfonts}
\usepackage{amsmath}
\usepackage{mathrsfs}
\usepackage{amssymb}
\usepackage{graphicx}
\usepackage{color}
\usepackage[english]{babel}
\usepackage[autolanguage]{numprint}
\usepackage{adjustbox}
\usepackage[breaklinks,colorlinks,citecolor=blue,unicode]{hyperref}

\makeatletter
\g@addto@macro\npstyleenglish{%
\npthousandthpartsep{}%
}%
\makeatother
\bibpunct{(}{)}{;}{a}{}{,}

\newcommand{\lyaffirst}{\text{Lyman-$\alpha$ (Ly$\alpha$)}}
\newcommand{\lyaf}{\text{Ly$\alpha$ forest}}
\newcommand{\lya}{\text{Ly$\alpha$}}
\newcommand{\HI}{\text{HI}}
\newcommand{\NHI}{$N_{\text{\HI}}$}
\newcommand{\NHIO}{$N_{\text{\HI},0}$}
\newcommand{\NHIt}{N_{\text{\HI}} }
\newcommand{\bvel}{$b$}
\newcommand{\cmtwo}{\text{cm}$^{-2}$}
\newcommand{\vpfit}{\texttt{VPFIT}}
\newcommand{\bndist}{$b$-$N_{\text{HI}}$ distribution}

\begin{document}
\title{A New Measurement of the Temperature Density Relation of the IGM From Voigt Profile Fitting}

\author{Hector Hiss\altaffilmark{*,1,2}, 
Michael Walther\altaffilmark{1,2,3},  
Joseph F. Hennawi\altaffilmark{1,3}, 
Jos\'{e} O\~{n}orbe\altaffilmark{1},\\ 
John M. O'Meara\altaffilmark{6}, 
Alberto Rorai\altaffilmark{1,4,5}
and Zarija Luki\'c\altaffilmark{7}} 
\email{*hiss@mpia.de}

\altaffiltext{1}{Max-Planck-Institut f\"ur Astronomie, K\"onigstuhl 17,
69117 Heidelberg, Germany}
 \altaffiltext{2}{International Max Planck Research School for Astronomy \&
 Cosmic Physics at the University of Heidelberg}
 \altaffiltext{3}{Physics Department, Broida Hall, University of California Santa Barbara, Santa Barbara, CA
93106, USA}
\altaffiltext{4}{Kavli Institute for Cosmology and Institute of Astronomy, Madingley Road, Cambridge CB3 0HA, UK}
\altaffiltext{5}{Institute of Astronomy, Madingley Road, Cambridge CB3 0HA, UK}
\altaffiltext{6}{
Saint Michael’s College, Department of Chemistry and Physics, One Winooski
Park, Colchester, VT 05439}
\altaffiltext{7}{LBNL, 1 Cyclotron Road, Berkeley, CA
94720, USA}

\begin{abstract}%
  We decompose the \lyaffirst{} forest of an extensive sample of 75
  high signal-to-noise ratio and high-resolution quasar spectra into a
  collection of Voigt profiles. Absorbers located near caustics in the
  peculiar velocity field have the smallest Doppler parameters,
  resulting in a low-$b$ cutoff in the \bndist{} set primarily by the thermal
  state of intergalactic medium (IGM).  We fit this cutoff as a
  function of redshift over the range $2.0\leq z \leq 3.4$, which
  allows us to measure the evolution of the IGM temperature-density
  ($T= T_0 (\rho\slash \rho_0)^{\gamma-1}$) relation parameters $T_0$ and
  $\gamma$. We calibrate our measurements against mock Ly$\alpha$ forest data, 
  generated using 26 hydrodynamic simulations with different thermal histories from the THERMAL suite, 
  also encompassing different values of the IGM pressure
  smoothing scale.  We adopt a forward-modeling approach and
  self-consistently apply the same algorithms to both data and
  simulations, propagating both statistical and modeling uncertainties
  via Monte Carlo.  The redshift evolution of $T_0$ ($\gamma$) shows a suggestive
  peak (dip) at $z=2.9$ ($z=3$).  Our measured evolution of $T_0$ and $\gamma$ are generally in good
  agreement with previous determinations in the literature.
  Both the peak in the evolution of $T_0$ at $z = 2.8$, as well as the high temperatures $T_0\simeq 15000-20000\,$K 
  that we observe at $2.4 < z < 3.4$, strongly suggest that a significant episode of heating occurred after the end 
  of \ion{H}{1} reionization, which was most likely the cosmic reionization of \ion{He}{2}.
\end{abstract}
\maketitle


\section{Introduction}
The evolution of the thermal state of the low density intergalactic medium (IGM) provides us with insight into the nature and
evolution of the bulk ($\gtrsim 90 \%$) of baryonic matter in the Universe \citep{meiksin1, mcquinn2016}. 
Of special interest are the thermal imprints of cosmic reionization processes that heated the IGM. 

The IGM is believed to have undergone two 
major reheating events. The first is the reionization of hydrogen (\ion{H}{1} $\rightarrow$ \ion{H}{2}), likely driven by 
galaxies \citep{Faucher-Giguere2008,Robertson2015} and/or quasars 
\citep[QSOs, ][]{Madau_qso, Khaire16},  
which should be completed by redshift $z\sim 6$ \citep{mcgreer1}. The standard picture is 
that helium is singly ionized
(\ion{He}{1} $\rightarrow$ \ion{He}{2}) during \ion{H}{1} reionization, and that the second ionization of
Helium (\ion{He}{2} $\rightarrow$ \ion{He}{3}) occurred later during a \ion{He}{2} reionization
phase transition powered by the harder radiation emitted by luminous QSOs.
This process is expected to be completed by a redshift of around 2.7 \citep{worseck1}. 
These reionization processes are expected to significantly alter the thermal evolution of the IGM.

Long after reionization events which heat the IGM, the thermal
properties of the bulk of the intergalactic gas 
are well described by a tight power law temperature-density relation
of the form $T = T_0 (\rho/\rho_0)^{\gamma-1}$ \citep{hui1,mcquinn16},
parametrized by the temperature $T_0$ at mean density $\rho_0$ and an index
$\gamma$. This relation comes about naturally 
when the gas is mainly heated by photoionization and cooled due to cosmic expansion. 
Therefore, the evolution of $T_0$ and $\gamma$ serves as a diagnostic tool to
understand reionization phase transitions.
Note that during or just after a reionization process, the gas experiences
temperature fluctuations \citep{daloisio2015} that 
cause this relation to experience scatter or even become multivalued \citep{mcquinn09, compostella1}.

Although predominantly photoionized, residual \ion{H}{1} in the diffuse IGM
gives rise to \lyaffirst{} absorption, ubiquitously 
observed toward distant background quasars. This so-called
\lyaf{} has been established as the premier probe of the IGM 
and cosmic structure at redshifts $z \lesssim 6$. 
In the literature, different approaches were used for measuring the 
parameters of the temperature-density relation of the IGM from 
\lyaf{} absorption. 
Studies of
the statistical properties of the absorption, such as the power-spectrum of the
transmitted flux \cite[e.g.][]{Zaldarriaga1,theuns1}, the
average local curvature \citep{becker1,boera1}, the flux probability
distribution function \cite[e.g.][]{bolton2008, viel1, lee1}
and wavelet decomposition
of the forest \cite[e.g.][]{lidz1,Garzilli2} aimed to constrain the thermal state of the IGM.

In this work, we follow the approach used by \citet{schaye3, ricotti1} and \citet{mcdonald1} that treats the 
\lyaf{} as a superposition of discrete absorption profiles. 
This method was suggested by \citet{haehnelt1, ricotti1} and \citet{bryan1} and is based 
on the idea that the distribution of Doppler parameters $b$, i.e. line broadening, of \lya{} absorption in the IGM at a given redshift, 
has a sharp cutoff at low values that can be connected to the thermal state of the IGM. 

Generally the Doppler parameter $b$ of an absorber is determined by the contributions from
its thermal state and kinematic properties. 
The thermal contribution consists of microscopic random thermal motions in the gas, or \textit{thermal broadening}, whereas the kinematic contribution, 
often referred to as \textit{turbulent broadening}, results from the peculiar velocities in the IGM as well as the differential Hubble flow
across the characteristic size of an absorbing cloud, which is set by
the so-called pressure smoothing scale $\lambda_P$ 
\citep{HuiGnedin1998,schaye2,Rorai2013,Kulkarni2015,rorai_science}. If we observe many absorption features, we will occasionally encounter lines
from gas clouds which have a line-of-sight velocity component near zero, 
i.e. the velocity field is close to a caustic \citep{mcdonald1}. 
As the broadening of these absorbers is dominated by the thermal contribution, 
this results in a thermal state dependent cutoff in the distribution of Doppler parameters. 
Note that this cutoff will be subject to scatter due to effects such as fluctuations in temperature and 
ionizing background. 
Assuming that the cutoff is primarily set by the thermal state of the gas, its position will be dependent 
on the gas density due to the temperature-density relation, or in observable terms, the absorption line column density \NHI{}. 
This in turn means that there is a correlation between the position of the lower cutoff in the 
distribution of Doppler parameters as a function of column densities (\bndist{}) 
and the thermal state of the gas. Measuring the position of this cutoff can thus reveal the underlying $T_0$ and $\gamma$.

Recently, a measurement of the thermal state of the IGM based on the cutoff of the \bndist{} 
was carried out by \citet{rudie1} using a 
sample of 15 high-quality QSO sightlines. Using the analytic relations
between the cutoff of the \bndist{} and the temperature-density relation derived
by \citet{schaye3}, \citet{rudie1} 
measured a temperature at mean-density $T_0 = 1.94 \times 10^4\,$K at $z\sim 2.4$ (in
a broad redshift bin spanning $2.0<z<2.8$), 
which was $\sim 9000 \,$K higher than the value implied by curvature 
measurements at $z=2.4$ by \citet{becker1}. 
This discrepancy motivated 
\citet{bolton1} to revisit this measurement. Using hydrodynamical simulations,
they calibrated the relationship between the \bndist{} cutoff
and the temperature-density relation.
Applying this updated
calibration to the \citet{rudie1} \bndist{} cutoff measurement, \citet{bolton1} determined
a lower temperature $T_0(z=2.4) = [1.00^{+0.32}_{-0.21}] \times 10^4\,$K 
that is consistent with \citet{becker1}, and argued that the much higher
temperature measured by \citet{rudie1} resulted from incorrect assumptions in the
calibration. 

In this work we study the \bndist{} of an extensive sample of 75 high quality
QSO spectra, which allows us to measure the redshift
evolution of $T_0$ and $\gamma$ over the redshift range $2.0\leq z \leq 3.4$
with a much finer binning $\delta z = 0.2$ than
previous work. At each redshift we use mock \lyaf{} data from 26 
hydrodynamic simulations with different thermal histories to calibrate the
relationship between the cutoff in the \bndist{} and the thermal parameters ($T_0$, $\gamma$, 
$\lambda_{P}$) governing the IGM. The \lyaf{} of both the data and the simulations
are decomposed into individual absorption lines using the Voigt-profile fitting algorithm \vpfit{} \citep{vpfit}, and we
adopt a forward-modeling approach whereby the
same algorithms are self-consistently applied
to both data and simulations.

This paper is structured as follows. We introduce our dataset, Voigt-profile, and cutoff fitting procedure in \S~\ref{sec:data}. 
An overview of our hydrodynamic simulations is given in \S~\ref{sec:simulations}, where we also introduce the 
THERMAL (Thermal History and Evolution in Reionization Models of Absorption Lines) suite. 
In \S~\ref{sec:calibration} we discuss how we calibrate our method by applying the same fitting procedures to simulated sightlines. 
Our final results on the evolution of the thermal state of the IGM at $2<z<3.4$ are presented and discussed in section \S~\ref{sec:measurements}. 
We summarize our results in section \S~\ref{sec:discussion}. 

\begin{figure}
 \plotone{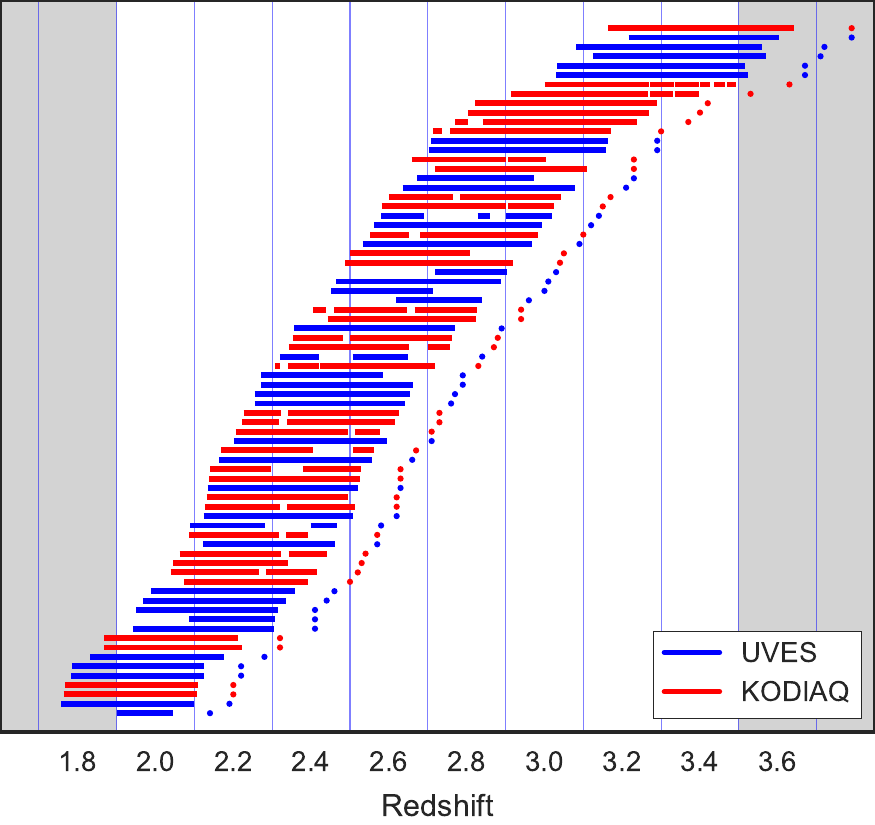}
  \caption{KODIAQ and UVES \lyaf{} sightlines used in this work. 
  The sample is described in \S~\ref{sec:data}. Sightlines from the KODIAQ sample are marked in red and UVES sightlines are marked in blue. The corresponding 
  QSO redshifts are marked as colored points.
  DLAs and bad regions that were excluded are shown as gaps. The blue vertical lines show the bins that will be used for the cutoff fitting analysis. 
  Gray regions are not used because of lower coverage.}
  \label{fig:qsosample}
\end{figure}

\section{Data Processing}
\label{sec:data}
\subsection{QSO Sample}
For this study we used a sample of 75 publicly available QSO spectra with signal-to-noise ratio (SNR) better than 20 per 6 km/s bin and resolution 
varying between ${\rm FWHM}=3.1$ km/s and 6.3 km/s with a typical value around 6 km/s. 
This ensures that the \lyaf{} is resolved and that we can detect lines with $\NHIt \simeq 10^{12.5}$ \cmtwo{} at the
3$\sigma$ level \citep{herbert-fort}. 
Part of the sample consists of QSO spectra from the 
Keck Observatory Database of Ionized Absorbers
toward QSOs \citep[KODIAQ,][]{kodiaq1,kodiaqdr1,kodiaqDR2}. The other
spectra were acquired 
with the UV-visual Echelle Spectrograph \citep{dekker} at the Very Large Telescope \citep{aldo1}.

The KODIAQ sample used in this work consists of 36 QSO sightlines chosen from DR1 and DR2. 
These QSOs were observed between 1995 and 2012 using the HIRES instrument (High Resolution Echelle Spectrometer: 
\citeauthor{hires} \citeyear{hires}) on the Keck-I telescope. All the spectra were 
uniformly reduced and continuum-fitted by eye by the KODIAQ team using the HIRedux 
code\footnote{HIRedux: \url{http://www.ucolick.org/~xavier/HIRedux/}}.
Spectra with multiple exposures were co-added in order 
to increase the SNR. The detailed information about the reduction steps is described in \citet{kodiaq2}.

The UVES spectra consist of 38 sightlines from the ESO
archive. These objects were chosen to have at least 10 exposures each
and complete (or nearly complete) \lyaf{} coverage. The data was
reduced by \citet{aldo1} using the \texttt{MIDAS} environment
ECHELLE/UVES and procedures described in \citet{kim1}. Each frame was
bias and background subtracted. Afterward, the echelle spectra were
extracted order by order assuming a Gaussian profile along the spatial
direction. The final co-added spectra have exquisite SNR 
per pixel $\geq40$ and a resolution of 6 km/s within the
\lyaf{}.
Continua were fitted by \citet{aldo1} using a cubic-spline
interpolation method. We used 38 spectra from the 40 available in this sample. 
One characteristic of the UVES pipeline is that the estimated errors at flux 
values close to zero is underestimated by a factor
of roughly two \citep{rdgen}. 
Therefore Voigt-profile
fitting algorithms will struggle to achieve a satisfactory $\chi^2$ for these
regions. To compensate for this, we used a dedicated tool implemented in
\texttt{RDGEN}\footnote{RDGEN:\url{http://www.ast.cam.ac.uk/~rfc/rdgen.html}} \citep{rdgen}, a front and 
back-end program for \vpfit{}. 
This tool multiplies the error of each pixel with a
value that is 1 if the corresponding normalized flux is 1 and 2 if the normalized flux is 0. 
For this purpose we used the default parametrization from \texttt{RDGEN}.

The region of the spectra used for 
fitting lies between 1050 \AA{} and 1180 \AA{} rest-frame inside the \lyaf{}. 
This region was chosen to 
avoid proximity effects, i.e. regions affected by the local QSO radiation rather than the metagalactic UV-Background.
This choice is consistent with studies by \citet{palanque} and \citet{Walther2017}. 

For a complete list of the spectra analyzed in this work and the essential information about them, 
refer to
Table \ref{table:qsotable}. The chunks of spectra used are plotted in Figure \ref{fig:qsosample} 
and colored based on the dataset they belong to. Our analysis of the thermal state of the IGM will be done in redshift 
bins of size $\delta z =0.2$, indicated with vertical blue lines. We discuss the effects of continuum 
misplacement in our data in the appendix~\ref{appendix}.

\begin{deluxetable}{llcc}
\tablecaption{QSO spectra used in this work. The signal-to-noise value refers to the median value inside the \lyaf{}. \label{table:qsotable}}
\tablehead{\colhead{Object ID} & \colhead{$z_{qso}$} & \colhead{$SNR / 6$ kms$^{-1}$} & \colhead{Sample}}
\startdata
HE1341-1020 & 2.137 & 58 & UVES \\
Q0122-380 & 2.192 & 56 & UVES \\
J122824+312837 & 2.2 & 87 & KODIAQ \\
J110610+640009 & 2.203 & 59 & KODIAQ \\
PKS1448-232 & 2.222 & 57 & UVES \\
PKS0237-23 & 2.224 & 102 & UVES \\
HE0001-2340 & 2.278 & 66 & UVES \\
J162645+642655 & 2.32 & 104 & KODIAQ \\
J141906+592312 & 2.321 & 37 & KODIAQ \\
Q0109-3518 & 2.406 & 70 & UVES \\
HE1122-1648 & 2.407 & 172 & UVES \\
HE2217-2818 & 2.414 & 94 & UVES \\
Q0329-385 & 2.437 & 58 & UVES \\
HE1158-1843 & 2.459 & 67 & UVES \\
J005814+011530 & 2.495 & 36 & KODIAQ \\
J162548+264658 & 2.518 & 44 & KODIAQ \\
J121117+042222 & 2.526 & 34 & KODIAQ \\
J101723-204658 & 2.545 & 70 & KODIAQ \\
Q2206-1958 & 2.567 & 75 & UVES \\
J234628+124859 & 2.573 & 75 & KODIAQ \\
Q1232+0815 & 2.575 & 46 & UVES \\
HE1347-2457 & 2.615 & 62 & UVES \\
J101155+294141 & 2.62 & 130 & KODIAQ \\
J082107+310751 & 2.625 & 64 & KODIAQ \\
HS1140+2711 & 2.628 & 89 & UVES \\
J121930+494052 & 2.633 & 90 & KODIAQ \\
J143500+535953 & 2.635 & 65 & KODIAQ \\
Q0453-423 & 2.663 & 78 & UVES \\
J144453+291905 & 2.669 & 134 & KODIAQ \\
PKS0329-255 & 2.705 & 48 & UVES \\
J081240+320808 & 2.712 & 49 & KODIAQ \\
J014516-094517A & 2.73 & 77 & KODIAQ \\
J170100+641209 & 2.735 & 82 & KODIAQ \\
Q1151+068 & 2.758 & 49 & UVES \\
Q0002-422 & 2.768 & 75 & UVES \\
HE0151-4326 & 2.787 & 98 & UVES \\
Q0913+0715 & 2.788 & 54 & UVES \\
J155152+191104 & 2.83 & 30 & KODIAQ \\
Q1409+095 & 2.843 & 25 & UVES \\
Q0119+1432 & 2.87 & 33 & KODIAQ \\
J012156+144820 & 2.87 & 55 & KODIAQ \\
Q0805+046 & 2.877 & 27 & KODIAQ \\
HE2347-4342 & 2.886 & 152 & UVES \\
J143316+313126 & 2.94 & 54 & KODIAQ \\
J134544+262506 & 2.941 & 35 & KODIAQ \\
Q1223+178 & 2.955 & 33 & UVES \\
Q0216+08 & 2.996 & 37 & UVES \\
HE2243-6031 & 3.011 & 119 & UVES \\
CTQ247 & 3.026 & 69 & UVES \\
J073621+651313 & 3.038 & 26 & KODIAQ \\
J194455+770552 & 3.051 & 30 & KODIAQ \\
HE0940-1050 & 3.089 & 70 & UVES \\
J120917+113830 & 3.105 & 31 & KODIAQ \\
Q0420-388 & 3.12 & 116 & UVES \\
CTQ460 & 3.141 & 41 & UVES \\
J114308+113830 & 3.146 & 32 & KODIAQ \\
J102009+104002 & 3.168 & 36 & KODIAQ \\
Q2139-4434 & 3.208 & 31 & UVES \\
Q0347-3819 & 3.229 & 84 & UVES \\
J1201+0116 & 3.233 & 30 & KODIAQ \\
J080117+521034 & 3.236 & 43 & KODIAQ \\
PKS2126-158 & 3.285 & 64 & UVES \\
Q1209+0919 & 3.291 & 30 & UVES \\
J095852+120245 & 3.298 & 45 & KODIAQ \\
J025905+001126 & 3.365 & 26 & KODIAQ \\
Q2355+0108 & 3.4 & 58 & KODIAQ \\
J173352+540030 & 3.425 & 57 & KODIAQ \\
J144516+095836 & 3.53 & 25 & KODIAQ \\
J142438+225600 & 3.63 & 29 & KODIAQ \\
Q0055-269 & 3.665 & 76 & UVES \\
Q1249-0159 & 3.668 & 70 & UVES \\
Q1621-0042 & 3.708 & 78 & UVES \\
Q1317-0507 & 3.719 & 42 & UVES \\
PKS2000-330 & 3.786 & 151 & UVES \\
J193957-100241 & 3.787 & 66 & KODIAQ
\enddata
\end{deluxetable}

\begin{figure*}
 \plotone{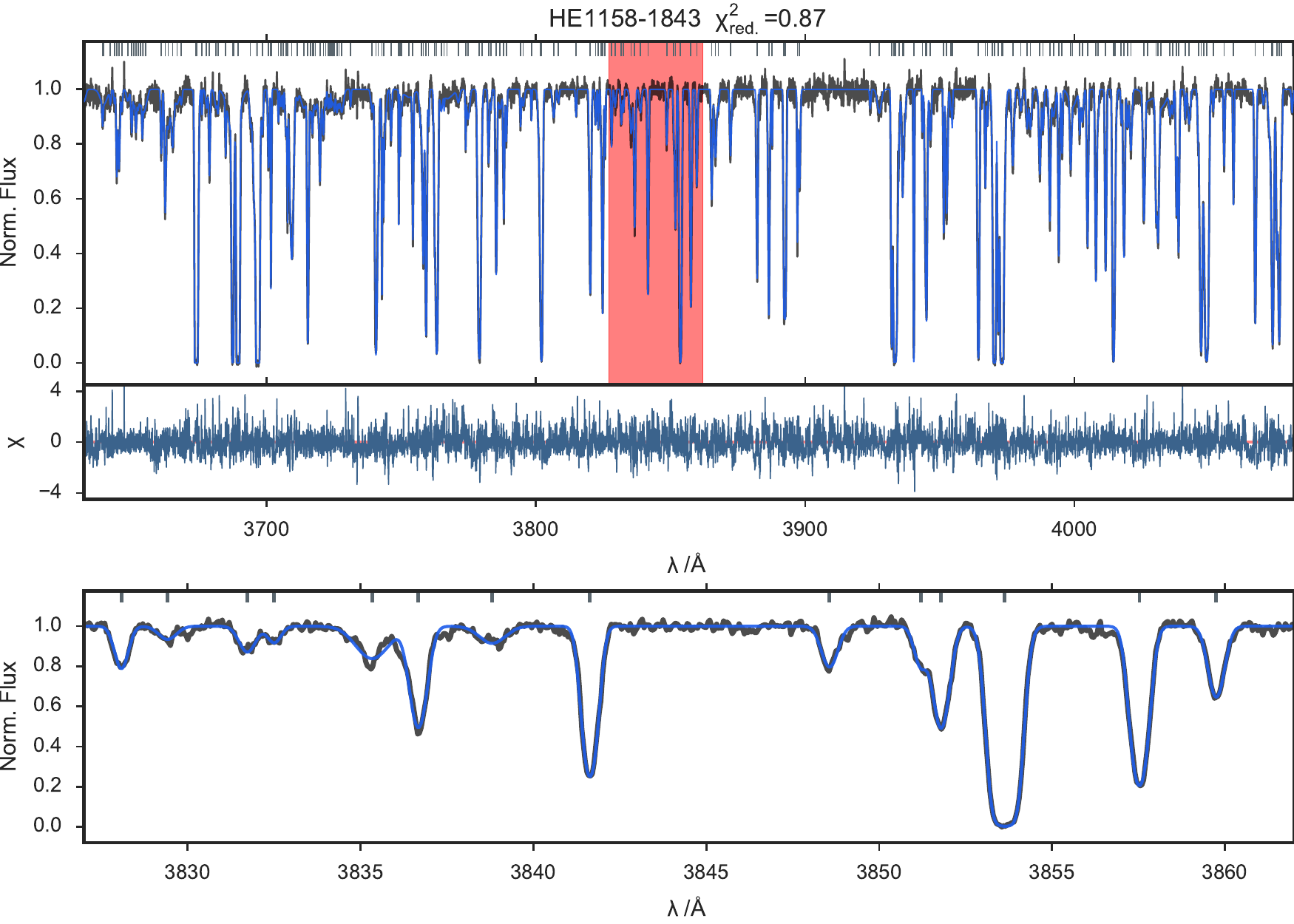}
  \caption[The resulting VP-fit of \lyaf{} the continuum normalized QSO HE1158-1843 at $z\simeq2.46$ from the UVES sample.]{The resulting VP-fit 
 of the \lyaf{} of the QSO HE1158-1843 at $z\simeq2.46$ from the UVES sample. \textbf{Upper panel}: The original spectrum (black line) is well described 
 by the superposition of Voigt-profiles fitted by \vpfit{} (blue line). The position of individual lines is shown by gray rugs in the upper part 
 of the panel. Underneath we plot the resulting $\chi =(F_{\text{spec}}-F_{\text{fit}})/
 \sigma_{F_{\text{fit}}}$ as a measure for the goodness of the fit. \textbf{Lower Panel:} Zoom in of the area marked in red 
 in the upper panel.}
 \label{fig:qsofit1}
\end{figure*}

\subsection{Voigt-Profile Fitting}
\label{sec:voigt-profile-fitting} 
Voigt-profiles are fitted to our data using \vpfit{} version 10.2\footnote{VPFIT: \url{http://www.ast.cam.ac.uk/~rfc/vpfit.html}} 
\citep{vpfit}.
We wrote a fully automated
set of wrapper routines that prepares the spectra for the fitting procedure and 
controls \vpfit{} with the help of the \vpfit{} front-end/back-end programs \texttt{RDGEN} and \texttt{AUTOVPIN}. These are used 
to generate initial guesses for the absorption line parameters, 
output tables, and determine which segments to fit separately.
For each segment \vpfit{} looks for the best fitting superposition of Voigt-profiles that describes a given spectrum. 
Each line is described by three parameters: line redshift $z_{\text{abs}}$, Doppler parameter $b$, and column density \NHI{} 
corresponding to the chosen absorbing gas transition (here hydrogen Ly-$\alpha$). 
The parameter space chosen for \vpfit{} to look for lines was set to go from 1 to 300 km/s in $b$ and 11.5 to 16.0 
in $\log(N_{\text{HI}}$/\cmtwo{}). 
\vpfit{} then varies these parameters and searches for a solution that minimizes the $\chi^2$. 
If the $\chi^2$ is not satisfying, then it will add lines until the fit converges or no longer improves. 
In order to minimize computational time, 
this fitting procedure is done in different segments of the spectra at a time. 
It is possible to automatically find regions that are 
between sections of the spectra where the flux meets the continuum, i.e. no absorption, and fit them separately. The fitted 
spectrum is then put together as a collection of line parameters.

Damped \lya{} systems (DLAs), i.e. \lya{} absorbers with $N_{\text{HI}}\gtrsim10^{20}$\cmtwo{}, were identified by eye and 
are excluded from our analysis. 
The DLAs were chosen to enclose a region between the two points where the damping wings reach the QSO continuum within the flux error.
Additionally, regions larger than 30 pixels previously masked in the data (bad pixels, gaps, etc.) were also 
excluded. We simply cut out the regions in which these rejections apply and feed the usable data segments into \vpfit{} separately.
 
In order to avoid chopping our spectra into too many small segments, small regions ($\leq$ 30 pixels) that were previously masked in the data 
were cubically interpolated. These pixels were given 
a flux error of a 100 times the continuum so the Voigt-profile fitting procedure is not influenced. 

One complication is that \vpfit{} often has difficulty fitting the
boundaries of spectra. To solve this problem we artificially make the
chunks longer. For this purpose we append a mirrored version of the first
quarter of the spectra to the beginning of it. We do the same
with the last quarter to the end of the spectrum. These regions and
the line fits within them are later ignored. This method ensures that
the unreliable fits at the boundaries happen in an artificial environment
that will not be used. The disadvantage is that the spectrum that
\vpfit{} receives is 50\% longer than the original and will therefore
need more time to be processed. 

An example of the VP-fitted spectrum of an UVES sightline is shown in Figure \ref{fig:qsofit1}.

\begin{figure}
 \plotone{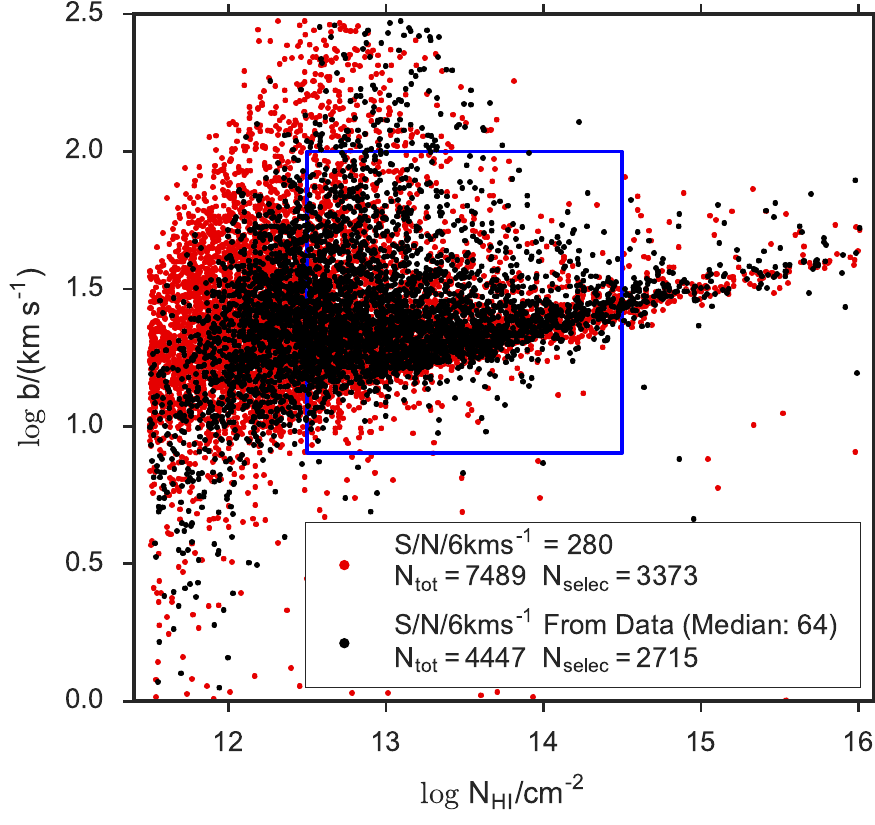}
  \caption{Simulated \bndist s at z=2.4 with different SNR 
  applied to lines-of-sight. The \bndist s were generated by VP-fitting the same 
  80 mock skewers from collisionless simulations, adding noise and resolution 
  effects. The \bndist{} based on high SNR skewers (red) has a higher number of fitted lines than the distribution 
  created based on the SNR distribution of our data at this redshift (black). 
  The high SNR distribution is more complete at low $\log N_{\text{HI}}$ high $\log b$. The blue box 
  shows the region chosen for our further analysis. The completeness is comparable within this box. 
  The thermal parameters used in these mocks are $\gamma=1.5$, $\log T_0/\text{K}=4.04$ and the smoothing length is 
  $\lambda_P = 47 \,$kpc.}
  \label{fig:SNtail}
\end{figure}

\subsection{The $b$-$N_{\text{HI}}$ Distribution}

The output of \vpfit{} can be used to generate a $\log b$ vs. $\log (N_{\text{HI}}/$\cmtwo) diagram ($b$-$N_{\text{HI}}$ distribution). 
Note that for a comparatively small number of lines, \vpfit{} outputs the errors as being 
zero, \texttt{nan} or ``*******''. When generating diagrams, we exclude these lines, because they normally appear in blended regions and 
noisy parts of the spectra. 

In order to illustrate the effect of SNR on the \bndist,
we generate 2 \bndist s by Voigt-profile fitting mock \lyaf{} absorption spectra at $z=2.4$ 
with different SNR applied to them. For this simple exercise we used mock \lyaf{} spectra based on 
collisionless dark matter only simulations\footnote{
These simulations use an updated version of the TreePM code from \citet{White02}, similarly to \citet{Rorai2013, rorai_science}, 
that evolves 
$N_p = 2048^3$ collisionless, equal mass particles ($M_p =2.5 \times 10^5 \text{M}_{\odot}$) in a periodic cube of side length
$L_{\text{box}} = 30\,\text{Mpc}/h$, adopting a \citet{planck} cosmology.}. 
The resulting distributions are shown in Figure \ref{fig:SNtail}. 
In one case (red) we added a constant and extremely high 
SNR/6\,kms$^{-1}$ of 280, while in the other case 
(black) a SNR based on the data at $z=2.4$ (with a median of SNR/6\,kms$^{-1}$=64) was applied. 
Some of the features are identical, especially the existence and position of a cutoff at 
$\log N_{\text{HI}}/\text{\cmtwo}>12.5$ and $\log b/$(kms$^{-1}$) $\sim 1.2$.
The main difference is that the high SNR distribution is more complete towards low $\log N_{\text{HI}}$ and 
high $\log b$ values. 
At column densities $\log N_{\text{HI}}/\text{\cmtwo}>12.5$ and Doppler parameters $8\, \text{km/s}\leq b \leq100\, \text{km/s}$ both 
distributions are similarly populated. 
Therefore for the cutoff fitting procedure we will only use the part of the \bndist{} with $\log N_{\text{HI}}/\text{\cmtwo}>12.5$, 
which is the convention adopted in
\citet{schaye1} and \citet{rudie1}. 
We also want to avoid saturated absorbers, i.e. $N_{\text{HI}}>10^{14.5}$\cmtwo, to make sure that we are using only well constrained 
column densities. 
Lines with $b<8$ km/s are excluded because these are most likely metal line contaminants or \vpfit{} artifacts. 
Lines with $b>100$ km/s are excluded as well, 
because the turbulent broadening component dominates over thermal broadening for such broad lines.
This is the same convention used in \citet{rudie1} and is shown as a blue box. 

Additionally, we decided, based on \citet{schaye3}, to exclude points that have relative errors worse than 50\% in \bvel{} or 
$N_{\text{HI}}$. This is done to avoid using badly constrained absorbers in the procedure, as they 
lie mostly in the part of the $b$-$N_{\text{HI}}$ distribution that is affected by the SNR effects 
described in Figure \ref{fig:SNtail} at high-\bvel{} and low-$N_{\text{HI}}$. 
Lines with $b<11$km/s, i.e. below the low-\bvel{} envelope of the distribution 
are generally not excluded by this procedure. 

\subsubsection{Metal masking}
\label{sec:metalmask}
It is well known that narrow absorption lines arising from ionic metal
line transitions contaminate the \lya{} forest, and will particularly impact the
lower $b\lesssim 10 \,$km/s region of the \bndist{} if treated as \lya{} absorption, 
thus possibly making the determination of the position of the lower envelope of the \bndist{} ambiguous.
To address this issue we remove lines from our sample that are potentially of metal origin.

However, narrow absorption lines are not
necessarily metal line contaminants. We visually inspected the
absorption lines with $b\lesssim10$km/s in every sightline and found that
although many could be identified as metal lines wrongly fit as \lya{} absorption, a comparable
number are simply narrow components that VPFIT adds to obtain
the best fit to complex Ly$\alpha$ absorption features. 
The latter are a property of the fitting procedure and should not be
excluded, as they are present in both data and the simulated spectra
that we use to conduct our 
analysis\footnote{For a discussion about how to circumvent the ambiguities associated with line deblending see \citet{mcdonald1}.}. 
In order to diminish the problem of metal line contamination we remove 
metal line contaminants combining automated and visual identification methods, which we
describe in detail below. 

Metals are typically associated
either with strong HI absorption, or they can be identified via associations with other
ionic metal line transitions.
Therefore, we identified 
DLAs based on the damping wings of the absorption profiles 
and determined their
redshifts with the help of associated metal absorption redwards
of the \lya{} emission peak of the QSO in question. 
The redshifts of other strong metal absorption systems 
not associated to a DLA within the data coverage or significantly shifted from a DLA
are determined by searching 
for typical doublet absorption systems (mostly \ion{Si}{4}, \ion{C}{4},
\ion{Mg}{2}, \ion{Al}{3}) redwards of the QSO's \lya{} emission peak. 
In both cases the doublets are identified based on their characteristic 
$\Delta \lambda$ (see Table~\ref{tab:absorbers}) and line-ratios.

\begin{figure}
 \plotone{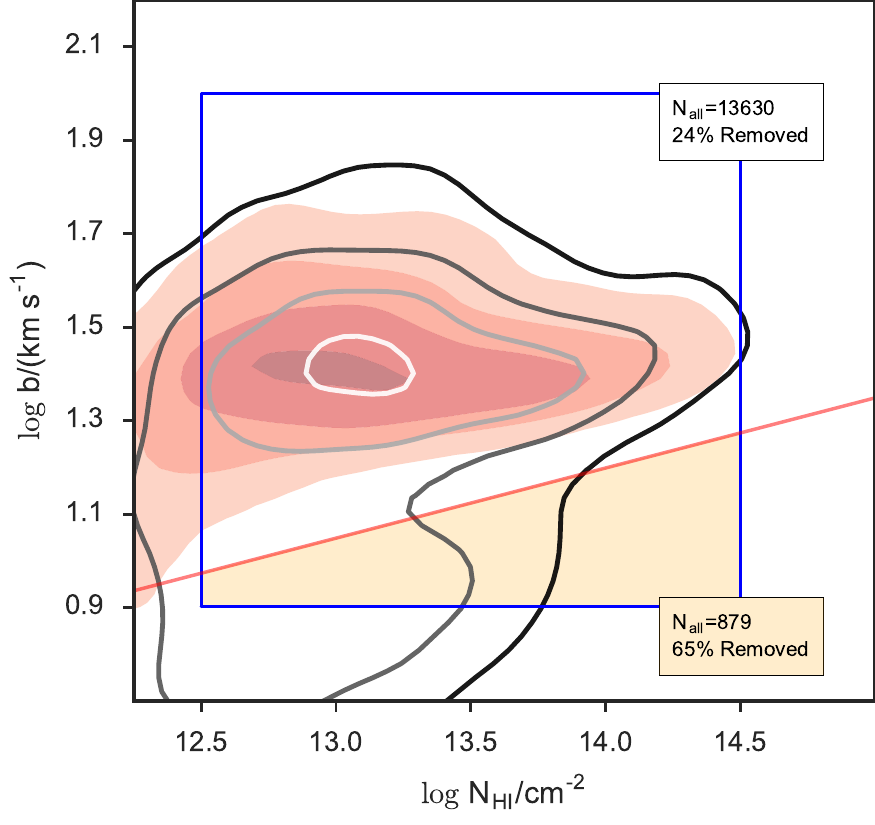}
 \caption{All lines in our QSO sample divided into two groups: the ones that were rejected using our narrow line rejection methods 
 (gray contour lines)
 and the ones that were kept and will be used for further analysis (red filled contours). For the sake of visibility, 
 we plot the two clouds of 
 points as continuous and normalized density distributions, calculated using a kernel density estimation method. The lines correspond to 
 5 equally spaced bins in density, i.e. the 80, 60, 40 and 20 percentiles of the corresponding density distributions.
 The blue square corresponds to our cutoff fitting region. 
 The solid red line broadly represents the dividing line between the bulks of the distributions of broad and narrow 
 lines with with $b < 11 \, \text{km/s} \times (N_{\text{HI}}/10^{12.95}{\text{cm}^{-2}})^{(1.15-1)}$. 
 To illustrate the region mostly affected by narrow lines in our cutoff fitting procedure, 
 we show the orange region. The fact that the red contours have little density below the red line, indicates that our 
 metal rejection 
 methods exclude most of the contamination. This happens at the cost of fraction of the usable data, i.e. the lines 
 in the gray contours that are not narrow. The total (both rejected and accepted together) number of lines $N_{\text{all}}$ 
 within the blue square 
 is shown above and below the solid red line, as well as the percentage of these lines that were 
 rejected as possible metal absorbers.}
 \label{fig:metal_rej}
\end{figure}

Additionally, we selected lines with Doppler parameters 
$b < 11 \, \text{km/s} \times (N_{\text{HI}}/10^{12.95}{\text{cm}^{-2}})^{(1.15-1)}$ in the \bndist s (red line in Figure \ref{fig:metal_rej}) 
and traced them back to their positions in the spectra. 
This relation was chosen based on visual inspection of the \bndist s at all redshift bins 
and chosen to lie underneath the lower envelope of the full dataset. 
We checked if we could find a match for 
different doublet ionic transitions within the \lyaf{} for these lines (typically \ion{Si}{4}, \ion{C}{4} and \ion{Mg}{2}) 
by testing for the $\Delta \lambda$ and line-ratios. We then confirmed 
them by finding corresponding absorption of other metals redwards of the \lya{} emission 
peak of the QSOs at the same redshift.
We then tested if the remaining lines below the lower envelope of the \bndist{} 
were any of the metal 
transitions listed in Table~\ref{tab:absorbers} by checking if other metal transitions 
and \lya{} absorption appear at the same redshift. The redshifts of systems positively 
identified as a metal line absorption with this method are stored. 
Candidate metal line absorbers only identified via a single metal feature or a doubtful
doublet feature, i.e. with one of the components possibly within a superposition of absorption features, 
were not considered as secure metal identification and thus are not masked. 
Given that it targets the absorbers found during the VP-fitting procedure, this method has the advantage that it allows us to identify metal absorbers within the \lyaf{} region. 

To further refine our metal line search, we used a semi-automated procedure
to identify high column density ($N_\textrm{HI}/\text{\cmtwo}\simeq$15) \ion{H}{1} absorbers
in our sample\footnote{This algorithm was written and tested by John O'Meara.} 
as these might also be associated with strong metal absorption. 
This algorithm identifies groups of pixels in a spectrum that have flux at the relative positions of 
Ly$\alpha$, $\beta$, $\gamma$ 
and higher orders (if available) within one sigma threshold of zero. 
The detected systems are then visually compared to theoretical line profiles of absorbers with 
$\log(N_\textrm{HI}/\text{\cmtwo})=15,~16,~17$ in Ly$\alpha$ and higher transitions up to Ly$\gamma$. 
If the absorption profile resembles that of a strong absorber, the redshift of the absorption system is saved. 
If the absorption was stronger than the $\log(N_\textrm{HI}/\text{\cmtwo})=15$ profile, then associated metals 
were masked (not the \ion{H}{1} absorption). 

Once we have the redshifts of the metal absorption systems, we create a mask based on the relative wavelength positions 
of the metal transitions listed in Table~\ref{tab:absorbers}. 
All listed transitions are used for generating masks, except for the systems 
identified with the automated method, i.e. the ones associated with $\log(N_\textrm{HI}/\text{\cmtwo}) \geq 15$. 
In this case we opted for a reduced list of strong ionic transitions (indicated in Table~\ref{tab:absorbers}). 
In case the position of any line from the \vpfit{} output falls within $\pm$30km/s from a potential metal line, 
it is removed from the line list. 
Additionally, Galactic CaII (3968\AA{}, 3933\AA{}) absorption was masked with a
$\pm 150\,$km/s window. 

Figure \ref{fig:metal_rej} shows normalized contours for all lines rejected using the narrow line rejection method described above 
(gray contour lines) and the lines that were kept (red filled contours) in our sample. 
We also show the fraction of points rejected in different regions of the \bndist{}. 
Our metal line filtering approach will inevitably also filter out lines that are genuine \lya{} absorption because 
of the window size of 30 km/s used in the narrow line rejection, 
removing 24\% of the absorbers that are not narrow.
This effect is visible in the overlap of rejected and accepted absorbers at $\log b >11\,$km/s. 
Nevertheless, we identified and removed $65\%$ of all absorbers in our dataset that are likely to be metal line contamination 
within our cutoff fitting range.

\begin{deluxetable}{clcl}
\tabletypesize{\footnotesize}
\tablecolumns{2}
\tablewidth{0pt}
\tablecaption{List of masked metal transitions. \label{tab:absorbers}}
\tablehead{\colhead{Absorber}&\colhead{$\lambda_\mathrm{rest}/\AA$}&\colhead{Absorber}&\colhead{$\lambda_\mathrm{rest}/\AA$}}
\startdata
\ion{O}{6}\tablenotemark{a}  & 1031.9261        & \ion{Si}{4}\tablenotemark{a} & 1402.770 \\
\ion{C}{2}         & 1036.3367        & \ion{Si}{2}        & 1526.7066 \\
\ion{O}{6}         & 1037.6167        & \ion{C}{4}\tablenotemark{a}  & 1548.195 \\
\ion{N}{2}         & 1083.990         & \ion{C}{4}\tablenotemark{a}  & 1550.770\\
\ion{Fe}{3}        & 1122.526         & \ion{Fe}{2}        & 1608.4511\\
\ion{Fe}{2}        & 1144.9379        & \ion{Al}{2}        & 1670.7874\\
\ion{Si}{2}        & 1190.4158        & \ion{Al}{3}        & 1854.7164\\
\ion{Si}{2}        & 1193.2897        & \ion{Al}{3}        & 1862.7895\\
\ion{N}{1}         & 1200.7098        & \ion{Fe}{2}        & 2344.214\\
\ion{Si}{3}\tablenotemark{a} & 1206.500         & \ion{Fe}{2}        & 2374.4612\\
\ion{N}{5}         & 1238.821         & \ion{Fe}{2}        & 2382.765\\
\ion{N}{5}         & 1242.804         & \ion{Fe}{2}        & 2586.6500\\
\ion{Si}{2}\tablenotemark{a} & 1260.4221        & \ion{Fe}{2}        & 2600.1729\\
\ion{O}{1}         & 1302.1685        & \ion{Mg}{2}        & 2796.352\\
\ion{Si}{2}        & 1304.3702        & \ion{Mg}{2}        & 2803.531\\
\ion{C}{2}         & 1334.5323        & \ion{Mg}{1}        & 2852.9642\\
\ion{C}{2}*        & 1335.7077        & \ion{Ca}{1}        & 3934.777\\
\ion{Si}{4}\tablenotemark{a} & 1393.755         & \ion{Ca}{1}        & 3969.591

\enddata
\tablenotetext{a}{Strongest transitions. The technique based on high density 
\lya{} systems filters only for these transitions.}
\end{deluxetable}

\subsubsection{Narrow Line Rejection}
\label{subsec:2sig}
Even after a careful metal line masking procedure, 
many unidentified narrow lines still remain in our line lists. These are 
narrow lines in blends and unidentified metal lines. 

One option to avoid these lines is by simultaneously fitting absorption 
profiles in the Lyman-$\beta$ (or higher transitions) forest, as in \citet{rudie1}. 
While this approach may deliver cleaner \bndist s, 
reproducing the same procedure applied to the data on simulations is very complicated as it
requires that one models higher-order Lyman series absorption as well. 
Furthermore,
the \citet{rudie1} selection of lines was not completely automated, and decisions about what lines
to keep were made by eye, which cannot be automatically applied to simulations \citep[see][for more details]{rudie2}. 
Therefore, in this work we chose to use only the \lyaf{} region.

Since there is no obvious way of filtering the remaining narrow lines, we need to come up with a rejection mechanism to filter them and diminish their 
impact on our cut-off fitting procedure.
To account for this problem \citet{schaye3} removed all the points in the \bndist{} where the best fitting 
Hui-Rutledge function\footnote{A one parameter function that describes the distribution of Doppler parameters 
$b$ under the assumptions that $\ln \tau$ is a Gaussian random variable, where $\tau$ is the optical depth, and that absorption lines 
arise from peaks in the optical depth \citep{HuiRutledge1999}.} to the $b$-distribution 
dropped below $10^{-4}$ at the low $b$ end. 
In \citet{rudie1}, the authors applied a more sophisticated algorithm that iteratively removes points from the $b$ distributions 
(with $b<40$ km/s) in $\log N_{\text{HI}}$ bins in case they are more than $2\sigma$ away from the mean. 

In this work we approach this problem in a very similar way as in \citet{rudie1}
Our rejection algorithm 
bins the points within $12.5 \leq \log(N_{\text{HI}}$/\cmtwo{})$\leq14.5$ into 6 bins of equal size in $\log(N_{\text{HI}}$/\cmtwo{}). 
Only points 
with \bvel{} $<$ 45 km/s
\footnote{The cut in $b<45$ km/s was chosen to be higher than the one used in \citet{rudie1}, because lower values were causing the rejection at $2\sigma$ 
to lie too close to the estimated position of the cutoff at some of the redshift bins. The higher cut in $b$ increases the dispersion per bin, making our 
rejection more conservative.} 
are used for the 2$\sigma$ rejection process. 
For each of the aforementioned 
bins we compute the mean and the variance of \bvel{}. Points below $2\sigma$ of the mean are excluded. 
This procedure is iterated until no points are excluded. 
Finally, after the last iteration,
we fit a line to the $\log b_{2\sigma}$ values of each $\log(N_{\text{HI}})$ bin. 
Once the position of this line is determined, 
we exclude all points below it from the original sample.
We have tested this algorithm for the effect of varying the $\sigma$ threshold and found that the end results 
are consistent with each other within the errors. 

\begin{figure}
 \plotone{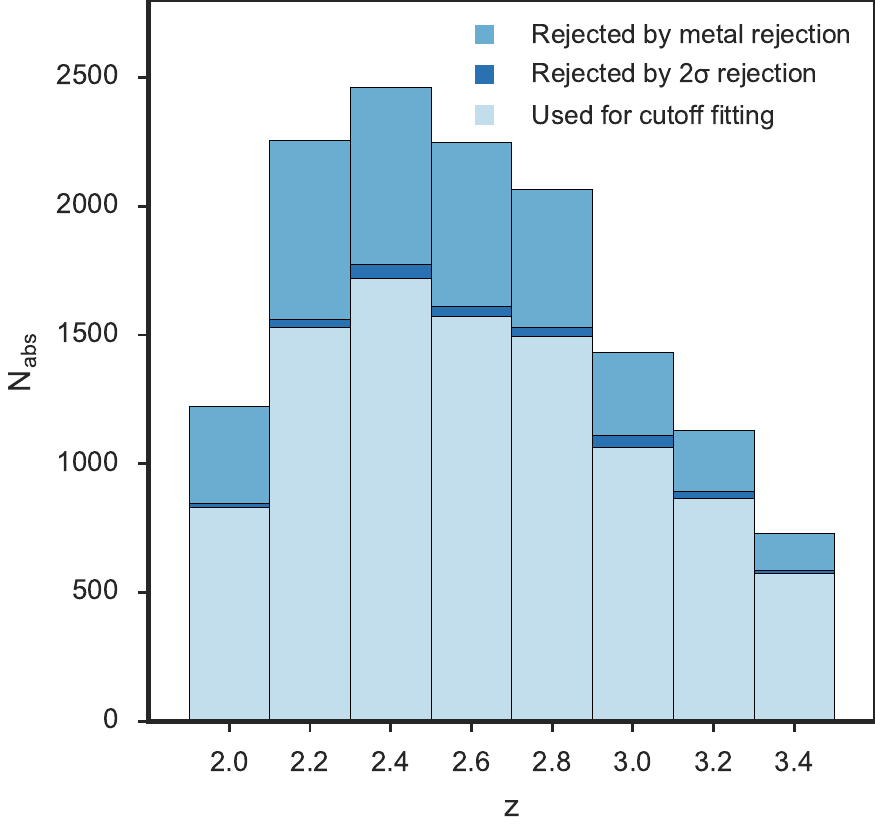}
 \caption{Number of absorbers fitted by {\tt VPFIT} per $\delta z$ bin. 
 The histogram shows the number of lines within the cutoff fitting range after metal lines rejection and the 2$\sigma$ 
 rejection were applied.} 
 \label{fig:absorbersdist}
\end{figure}

In Figure \ref{fig:absorbersdist} we show a histogram with the number of absorbers in every 
redshift bin of our data sample and the effects of rejections. Here we see that the $2\sigma$ rejection excludes a relatively small fraction 
of the points in the \bndist{}. 

\begin{figure*}[ht]
  \begin{adjustbox}{addcode={\begin{minipage}{\width}}{\caption{%
      The \bndist s in the redshift range $1.9 \leq z < 3.5$ in $\delta z = 0.2$ bins (corresponding to the sightlines in 
      Figure \ref{fig:qsosample}). 
      The best cutoff fits (red) and 2$\sigma$-rejection (black) lines are overplotted. 
      The shaded blue region represents the 68\% confidence region of the fits to bootstrap realizations at 
      every column density. The corresponding $N_{\text{HI,0}}$ is plotted as an open red point 
      and is calculated by plugging in the bin center redshift into eqn.~\ref{eq:NHI0}.
      These measurements allow us to access the evolution of $b_0$ and $\Gamma$ as a function of redshift.
      } \label{fig:VPmatrix}\end{minipage}},rotate=90,center}
      \epsscale{1.2}
      \includegraphics[scale=0.8]{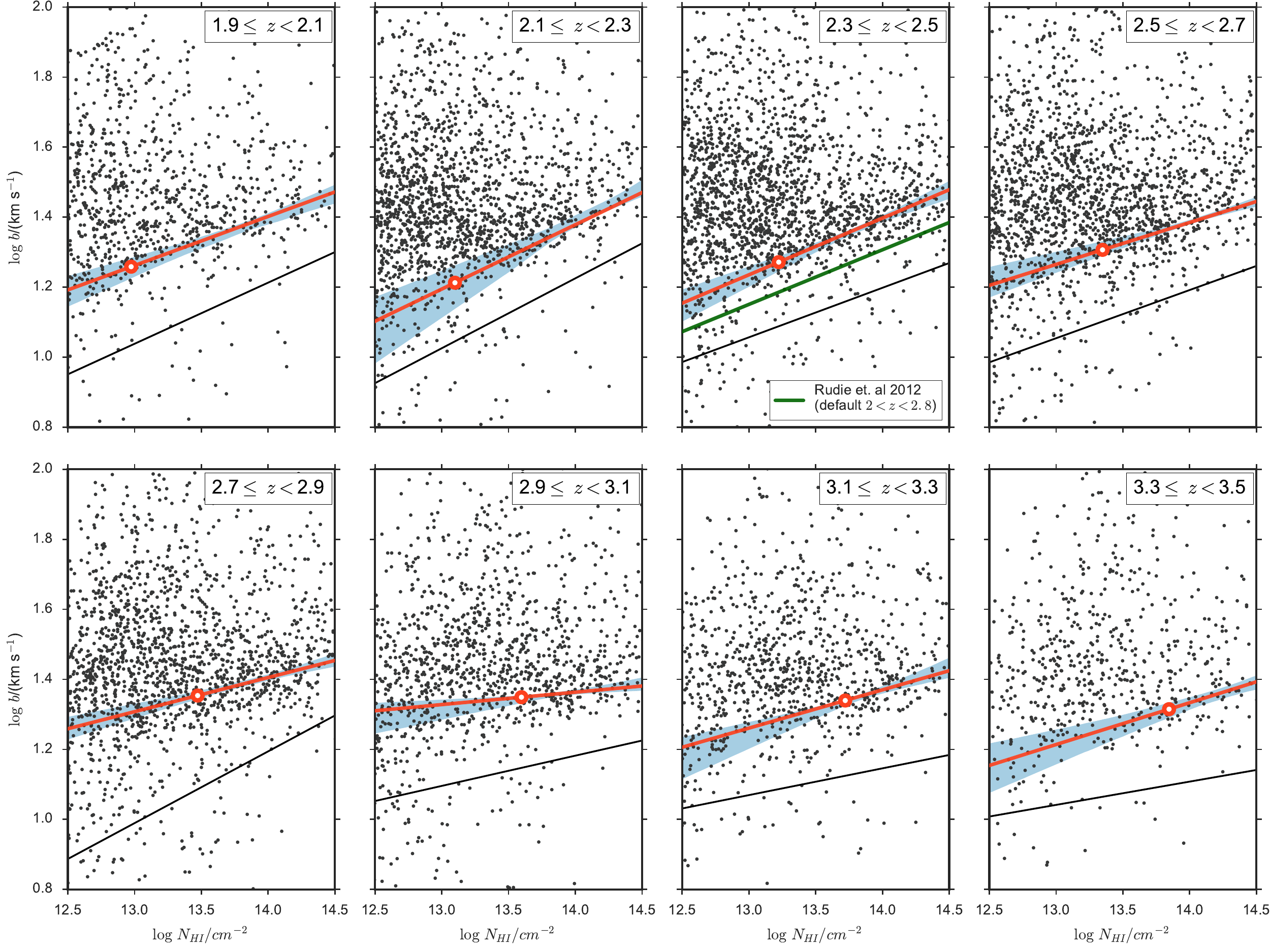}
  \end{adjustbox}
\end{figure*}
\begin{figure*}
\plotone{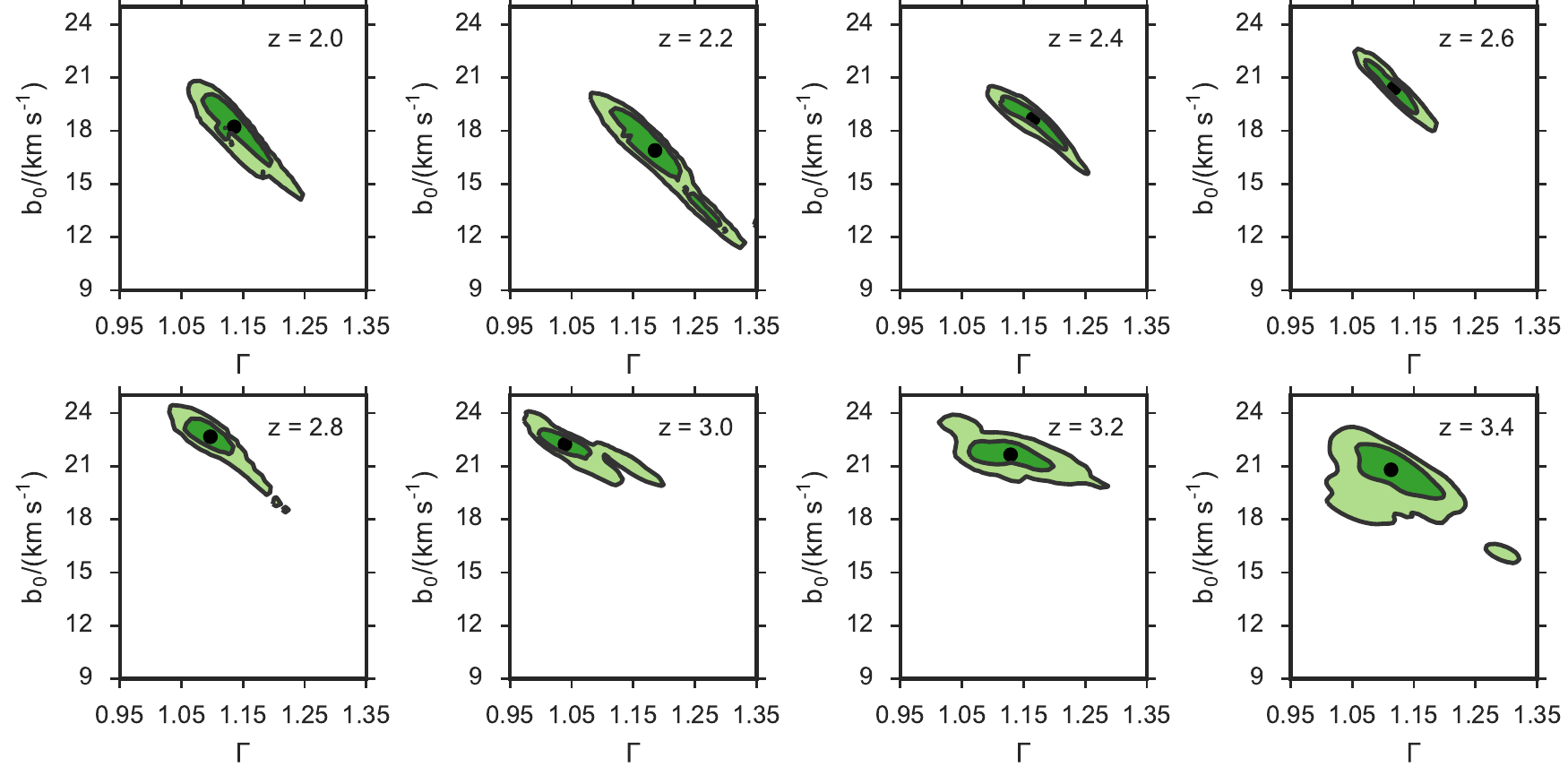}
\caption{Resulting $p(b_0,\Gamma)$ pdfs generated from cutoff fits to 2000 bootstrap realizations of the \bndist s at each redshift. 
The 68\% confidence levels are plotted in dark green and 95\% in light green. The black point corresponds to the median of 
the marginal distributions of $b_0$ and $\Gamma$.} 
\label{fig:b0Gammapdf}
\end{figure*}
\subsubsection{Fitting the Cutoff in the $b$-$N_{\rm HI}$ Distribution}
\label{sec:iterativefit}
Once we have the \bndist s, we want to determine where the thermal state sensitive cutoff is positioned.
The position of the cutoff
is calculated using our version of an iterative fitting procedure first introduced by \citet{schaye3} and 
also used in \citet{rudie1}. 
The function used for the cutoff of the $b$-$N_{\text{HI}}$ distribution is given by 
\begin{equation}
  \log b_{\text{th}} = \log b_0 + (\Gamma-1) \log (N_{\text{HI}}/N_{\text{HI,0}}).
 \label{eq:fit_func}
\end{equation}
where $b_0$ is the minimal broadening value at column density $N_{\text{HI,0}}$ and $\Gamma$ 
is the index of this power law relation.

Although the value of $N_{\text{HI,0}}$ is essentially just a normalization, as we will
motivate further in our discussion of the estimation of $N_{\text{HI,0}}$ in \S~\ref{sec:NHI0},
it is convenient to choose it
so that it corresponds to the column density of a typical absorber at the mean
density of the IGM. \citet{schaye2} showed that an absorber corresponding to an overdensity $\Delta = \rho/\rho_0$ with size of order of the 
IGM Jeans scale will have a column density
\begin{equation}
\label{eq:boltonNHI}
 N_{\text{HI}} \simeq 10^{13.23} \text{cm}^{-2} \Delta^{3/2} \frac{T_4^{-0.22}}{\Gamma_{\text{ion,HI}}} \left( \frac{1+z}{3.4} \right)^{9/2},
\end{equation}
where $\Gamma_{\text{ion,HI}}$ is the photoionization rate of \ion{H}{1} and $T_4$ is the temperature of the absorbing gas in 
units of $10^4\,K$. 
We compute $N_{\text{HI,0}} = N_{\rm HI}(\Delta=1)$ at each redshift using this eqn. and discuss
how it impacts our calibration in \S~\ref{sec:NHI0}.

In our iterative cutoff fitting procedure, we fit eqn.~(\ref{eq:fit_func}) to points in the \bndist{} 
using a least-squares minimization algorithm which takes into account the errors reported by \vpfit{}. 
Note that previous works \citep{schaye3, bolton1, Rorai17vpfit} have used a least absolute deviation method for fitting. 
For a method comparison and discussion see Appendix~\ref{app:alberto}.

The first step of the cutoff fitting procedure is to fit eqn.~(\ref{eq:fit_func}) 
to all points that are within 
$10^{12.5}\text{\cmtwo{}} < N_{\text{HI}} <  10^{14.5}\text{\cmtwo{}}$ and $8 \ \text{km/s}< b < 100 \ \text{km/s}$. 
The first iteration results in a fit that 
falls somewhere close to the mean of the distribution. Then we compute the mean absolute 
deviation in terms of $\log b$ of all $N$ absorbers with respect to the first fit:
\begin{equation}
\left< \left| \delta \log b \right| \right> = \frac{1}{N}\sum^{N}_i |\log b_{i}- \log b_{\text{th}}(N_{\HI,i})|.
\end{equation}
Notice that this takes the deviations both above and below the fit
into account. All the points that have a Doppler parameter with $\log
b> \log b_{\text{th}} +\left< \left| \delta \log b \right| \right> $
are excluded in the next iteration. 
This process is repeated without the points excluded in the previous iteration 
until no points are more than one absolute mean deviation above the fit, 
which defines convergence.
After convergence, the absorbers 
that are more than one mean deviation below the last fit 
are excluded. 
The remaining points are used for the final fit.

\subsection{Data cutoff fitting results}
\label{sec:data_cutoff}

Figure \ref{fig:VPmatrix} shows the $\log b(z)$-$ \log \NHIt{}(z)$ distributions resulting from the 
VP-fitting procedure and the respective cutoff fits (red) and 2$\sigma$ rejection lines (black). 
The values of $N_{\text{HI,0}}$ chosen for each cutoff fit are calculated using eqn.~(\ref{eq:NHI0}) 
at the central redshift of each bin. 
Their values are plotted as open red circles. 
We determine the uncertainty in the cutoff fit parameters via a bootstrap procedure. 
For this purpose, we generate the PDF $p(b_0,\Gamma)$ by bootstrapping the cutoff fitting procedure 2000 times using random realizations of the 
\bndist{} points with replacement. This results in a list with 2000 pairs of $(b_0,\Gamma)$. 
The 68\% confidence region of the bootstrap cutoff fits is shown in light blue. 
For illustration, a kernel density estimation of $p(b_0,\Gamma)$ at every redshift is shown in Figure \ref{fig:b0Gammapdf}. The anti-correlation 
between $b_0$ and $\Gamma$ is evident.

\section{Simulations}
\label{sec:simulations}
In this section we describe how we generate \lyaf{} mock spectra from Nyx hydrodynamic simulations 
\citep{nyx, Lukic2015} with different combinations 
of underlying thermal parameters $T_0$, $\gamma$ and $\lambda_P$. We apply the exact 
same Voigt-profile and \bndist{} cutoff fitting algorithms as for the data in order 
to calibrate the relations between the parameters that describe the cutoff ($b_0$ and $\Gamma$) 
and the thermal parameters ($T_0$ and $\gamma$) while marginalizing over different values 
of the pressure smoothing scale $\lambda_P$.

The evolution of dark matter in Nyx is calculated by treating dark matter particles as self gravitating Lagrangian particles, 
while baryons are treated as an ideal gas on a uniform Cartesian grid.
Nyx uses a second-order accurate piecewise parabolic method (PPM) to solve for the Eulerian gas dynamics equations, which accurately 
captures shock waves. For more details on the numerical methods and scaling behavior tests, see \citet{nyx} and 
\citet{Lukic2015}.
These simulations also include the physical processes needed to model the \lyaf{}. 
The gas is assumed to be of primordial composition with Hydrogen and Helium contributing 75\% and 25\% by mass. 
All relevant atomic cooling processes, as well as UV photo-heating, are modeled under assumption of ionization equilibrium.
Inverse Compton cooling off the microwave background is also taken into account. 
We used the updated recombination, cooling, collision ionization and dielectric recombination rates from \citet{Lukic2015}.

As is standard in hydrodynamical simulations that model the \lyaf{} forest,
all cells are assumed to be optically thin to radiation. Radiative feedback
is accounted for via a spatially uniform, but time-varying
ultraviolet background (UVB) radiation field,
input to the code as a list of photoionization and photoheating rates that
vary with redshift \citep[e.g.][]{Katz1992}. 
We have created a grid of models that explore very different thermal histories
combining different methodologies.
First we have used the approach presented in
\citet{jose1}, which allows us to vary the timing and duration of
reionization, and its associated heat injection, enabling us to simulate a
diverse range of reionization histories. This method allows us to create the
\ion{H}{1}, \ion{He}{1} and \ion{He}{2} photoionization and photoheating rates, which are
inputs to the Nyx code, by volume averaging the photoionization and energy
equations. We direct the reader to
\citet{jose1} for the details of this method.
On top of this we also use the methodology first introduced by
\citet{bryan1} 
of rescaling the photoheating rates by factor, $A$,
as well as making the heating depend on density according to $\Delta^{B}$ \citep{becker1},
with $B$ being also a free parameter.
Combining all these approaches allows us to built a large set of different 
thermal histories and widely explore the thermal parameter space of 
$T_0$, $\gamma$ and $\lambda_P$ at different redshifts.

The THERMAL\footnote{Url: \url{thermal.joseonorbe.com}} Suite (Thermal History and Evolution in Reionization Models of Absorption Lines)
consists on more than 60 Nyx hydrodynamical simulations 
with different thermal histories and $L_{\text{box}} = 20\,\text{Mpc}/h$ 
and $1024^3$ cells based on a \citet{planck} 
cosmology $\Omega_m=0.3192$, $\Omega_{\Lambda}=0.6808$, $\Omega_b =0.04964$, $h=0.6704$, $n_s=0.96$, $\sigma_8=0.826$. 
As shown in \citet{Lukic2015} for a \citet{HM} model, simulations of this box size and larger ones result 
in nearly the same distribution of column densities and Doppler parameters for the range 
of these parameters used in this work. 
The suite also has some extra simulations with different cosmological seeds, box size, 
resolution elements and/or cosmology to provide a reliable test bench for convergence and systematics
associated with different observables.
For all simulations we have data for every $\Delta z=0.2$ from $z=6.0$ down to $z=1.6$,
as well as at $z=1.0$, $z=0.5$ and $z=0.2$.

In this work we use a subset of 26 simulations from the THERMAL Suite that were selected to optimize
the space of thermal parameters (described below) within the redshift range in which we are
interested $2.0<z<3.4$.
The thermal parameters $T_0$ and $\gamma$ are extracted from the simulations by 
fitting a power law $T$-$\rho$ relation to 
the distribution of gas cells as described in \citet{Lukic2015}. 
In order to determine the pressure smoothing scale $\lambda_P$, the cutoff in the 
power spectrum of the real-space \lya{} flux 
$F_\mathrm{real}$ is fitted. $F_\mathrm{real}$ is the flux each position in the simulation 
would have given it's temperature and density, but neglecting redshift space effects \citep[see][]{Kulkarni2015}.

\subsection{Skewer Generation}
\label{subsec:skewers}
In order to model lines-of-sight through the IGM, we extract a random subset of hydrogen density skewers from our 
simulations that run 
parallel to the box axes. These are transformed into \lya{} optical depth skewers (we refer to \citealt{Lukic2015} for specific 
details about these calculations). 
The corresponding flux skewer $F$, i.e. a transmission spectrum along the line-of-sight, is calculated 
from the optical depth using $F = \exp(-A_r \, \tau)$. Here we introduce a scaling factor $A_r$ that allows us to match our 
lines-of-sight to observed mean flux values. 
This re-scaling of the optical depth accounts for the lack of knowledge of the precise value
of the metagalactic ionizing background photoionization rate. 
To this end we choose $A_r$ so that we match the mean-flux evolution shown in \citet{jose1}, which is a fit 
at $0.2<z<5.85$ based on measurements 
of various authors \citep{Fan06, Kim07,Faucher-Giguere2008, Becker_A}.
Given the extremely high precision with which the mean flux
has been measured by these authors, we do not consider the impact of uncertainties in the re-scaling 
value $A_r$. 
A discussion about the effects of mean flux rescaling in the models on our results is 
presented in the Appendix~\ref{app:mean_flux}.

\begin{figure}
\plotone{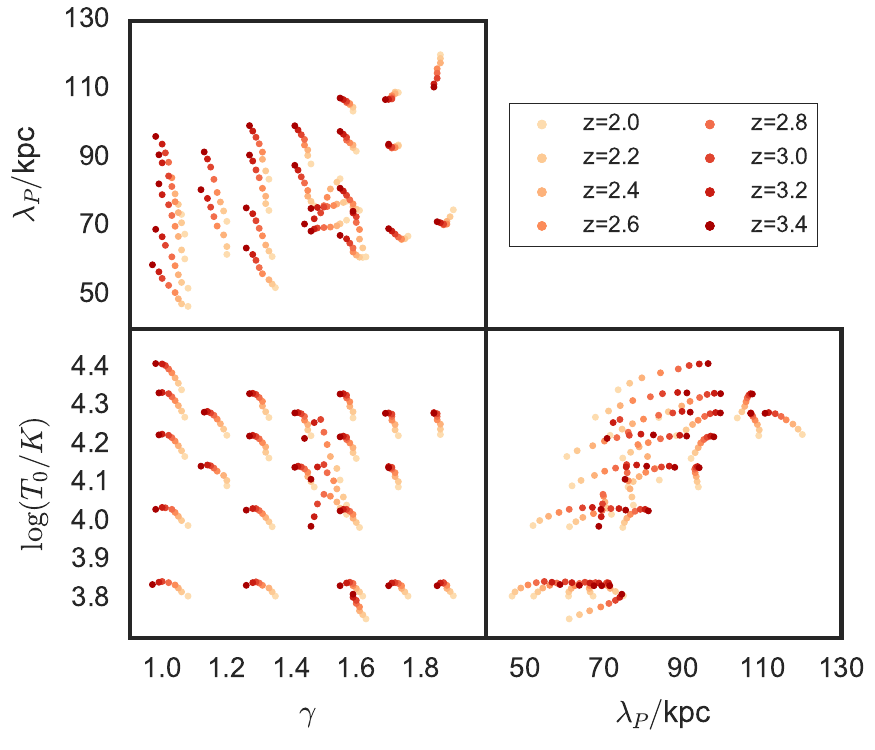}
\caption{Combinations of parameters $T_0$, $\gamma$ and $\lambda_P$ used for generating model skewers used in the calibration process. This grid was 
generated at redshifts 2.0 to 3.4 in $\delta z =  0.2$ steps to match our data. The evolution of the grid with redshift reflects 
the thermal history of the Nyx simulations chosen.} 
\label{fig:lamJdist}
\end{figure}
\subsection{Thermal Parameter Grid}
We used simulation snapshots at 8 different redshifts 
from $z=2.0$ to $3.4$ in $\delta z =0.2$ steps, 
which matches the redshift distribution of our data. 
We then generate 150 skewers for $2.0\leq z \leq 3.0$ and 75 skewers for $3.2 \leq z \leq 3.4$\footnote{These numbers of 
skewers were chosen based on the computation time needed for Voigt-profile fitting $z>3.2$ mock spectra at high SNR. 
Adopting these number results in nearly the same amount of absorbers in the \bndist{} used for cutoff fitting as in our data bins 
from $z=2$ to $2.6$ and $\sim 2500$ absorbers from $z=2.8$ to $3.4$.}
for each of the 26 combinations of thermal parameters ($T_0$, $\gamma$ and $\lambda_P$). 
Figure~\ref{fig:lamJdist} shows the distribution of thermal parameters chosen. 
We chose to model the thermal parameters on an irregular grid covering the range
$47 \,\text{kpc} < \lambda_P < 120 \,\text{kpc}$, which is well within the range of measurements by 
\citet{Rorai2013, rorai_science} of $40 \,\text{kpc} < \lambda_P < 130 \,\text{kpc}$ for $2<z<3.6$. For this comparison we 
scaled the measurements of \citet{Rorai2013, rorai_science} to match $\lambda_P$ as defined in \citet{Kulkarni2015}. 
The grid of parameters of the temperature-density relation covers $0.97 < \gamma < 1.9$ and $5600\ \text{K} < T_0 < 25700\ \text{K}$.

\subsection{Forward Modeling Noise and Resolution}
To create mock spectra we add the effects of resolution and noise, both based on our data, to our
simulated skewers. 
We mimic instrumental resolution by convolving the skewers 
with a Gaussian with FWHM = 6 km/s which is our typical spectral resolution
and re-binning to 3 km/s pixels afterwards. 
To make our mock spectra comparable to the data we added noise to the flux
based on the error distribution as provided by the data reduction pipelines. 
First, a random \lyaf{} at the same redshift interval is chosen from our QSO sample. 
A Gaussian pdf is constructed based on the median 
and a rank-based estimate of the standard deviation of the error distribution of the chosen data segment. 
Then, for every pixel $i$ in the skewer, we draw random errors $\epsilon_i$ from this pdf. 
We re-scale the errors so that $\epsilon_{i,r} = \sqrt{\Delta \lambda_{\text{data}}/ \Delta \lambda_{\text{skewer}}} \times \epsilon_i$, 
where $\Delta \lambda$ is the median wavelength distance between pixels. 
This accounts for the difference in sampling between data and skewers.
Finally we add a random deviate to the flux $F_i$ 
drawn from a normal distribution with $\sigma = \epsilon_{i,r}$, which is the error bar attributed to the flux. 
We do not account for metal line contaminants in our mock spectra, as these are explicitly masked in our data (see \S~\ref{sec:metalmask}). 

\begin{figure}
 \plotone{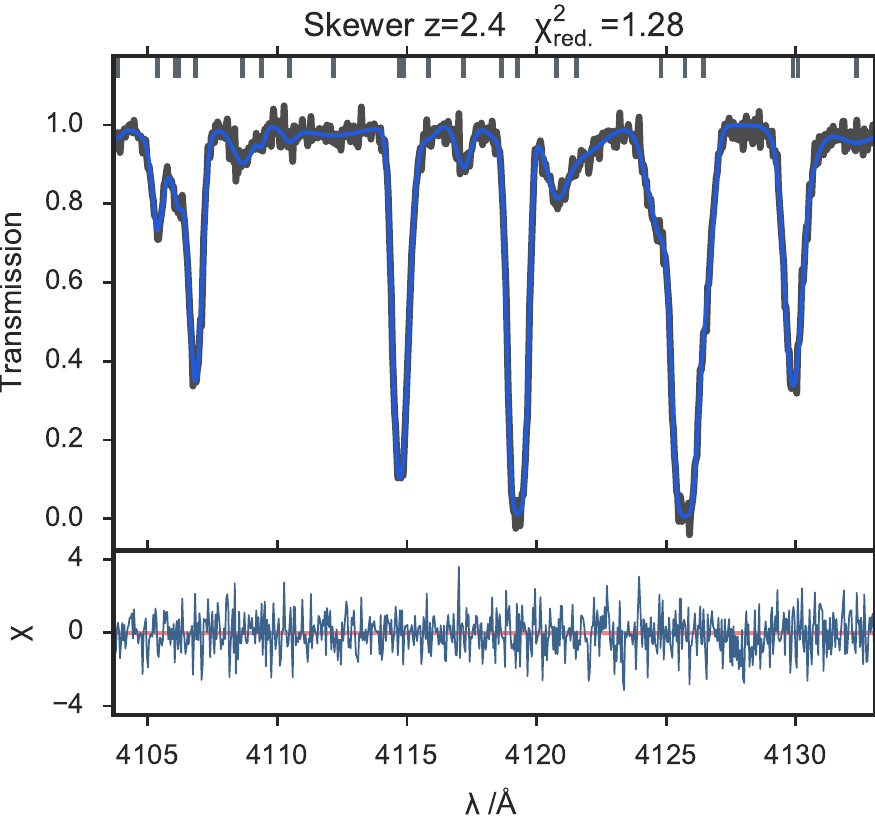}
  \caption{A VP-fitted simulated line-of-sight at z=2.4. 
The blue line is the spectrum fitted by \vpfit{}. The underlying 
black line is the original skewer. Noise was generated based on our data to achieve a SNR of 52 per pixel 
at continuum level. The simulation used had a best fit temperature-density 
relation with $\gamma$=1.52, $\log T_0/\text{K} = 4.07$ and a smoothing scale of $\lambda_P = 70 \, \text{kpc}$}. 
Underneath we we plot the resulting $\chi =(F_{\text{spec}}-F_{\text{fit}})/\sigma_{F_{\text{fit}}}$.
 \label{fig:skewerfit}
\end{figure}

\subsection{VP-fitting Simulations}
\begin{figure*}
 \plotone{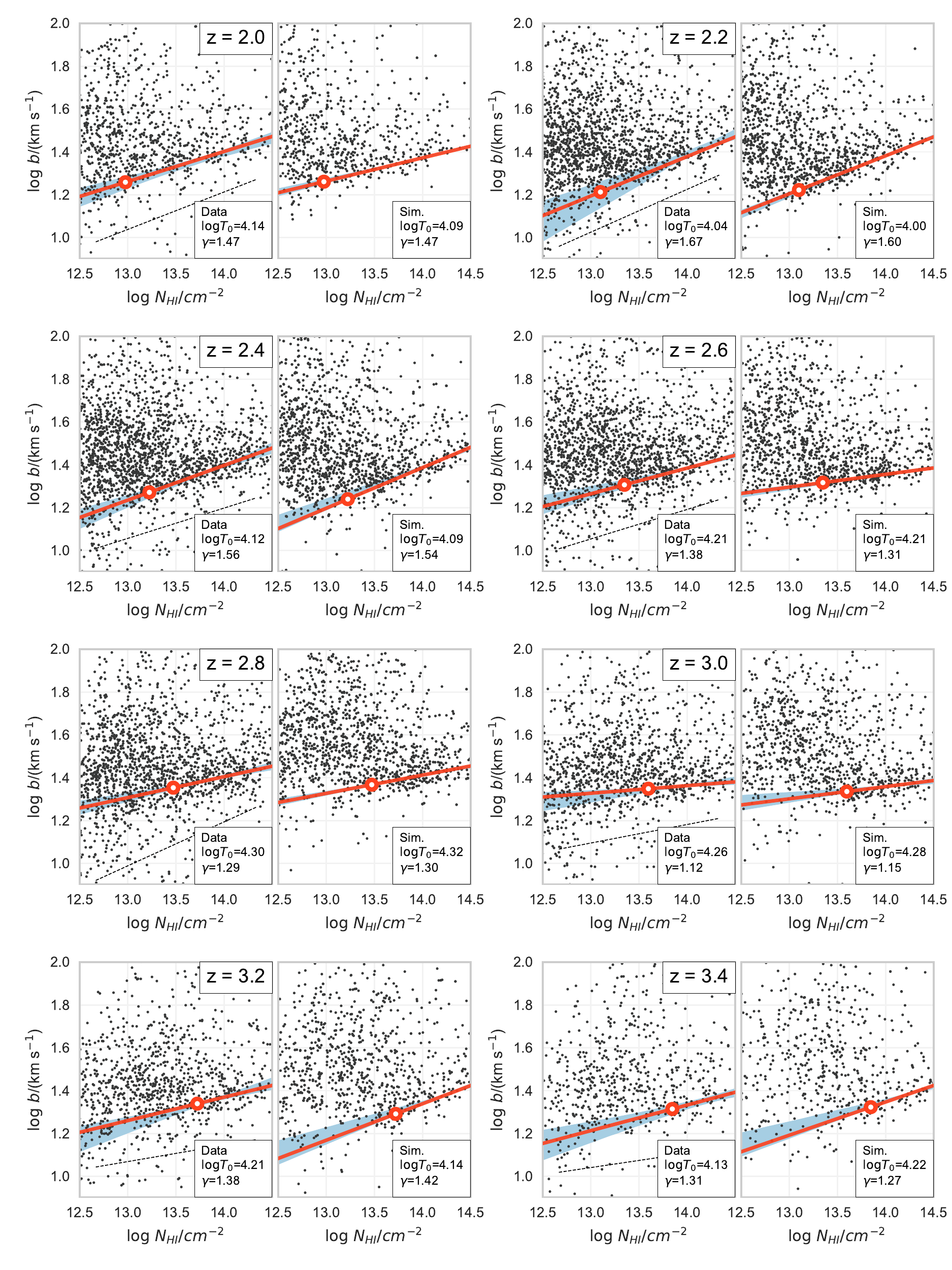}
 \caption{Comparison of \bndist{}s and cutoff fits for every redshift bin. 
 At each redshift our data are shown in the left panel, while the simulated \bndist{} is shown in the right panel.
 The simulated \bndist{}s shown are the ones that have $T_0$ and $\gamma$ closest to our final results (which will be introduced 
 in \S~\ref{sec:measurements}). 
 The best cutoff fits (red) and 2$\sigma$-rejection (black dashed, data only) lines are overplotted. 
 The shaded blue region represents the 68\% confidence region of the fits to bootstrap realizations at 
 every column density. 
 The value of $N_{\text{HI,0}}(z)$ is plotted as an open red point (the choice of $N_{\text{HI,0}}$ is motivated in \S~\ref{sec:NHI0}).
 Noise and resolution effects were added to skewers 
 based on the properties of our data. The cutoff fitting algorithm responds similarly 
 to data and models once the contamination in the data is removed using the 2$\sigma$ rejection algorithm. The remaining contamination in the data is still more severe 
 than in the models. This affects how the cutoff fitting procedure reacts to different bootstrap realizations.}
 \label{fig:bnsim}
\end{figure*}
We apply the exact same Voigt-profile fitting scheme described in \S~\ref{sec:voigt-profile-fitting} to the forward modeled 
simulated skewers 
generated for different combinations of $T_0$, $\gamma$ and filtering scale $\lambda_P$. 
A Voigt-profile fit of a mock spectrum is shown in Figure \ref{fig:skewerfit}. 
We then generate a $b$-$N_{\text{HI}}$ distribution for all our models and apply 
the same cutoff fitting algorithm described in \S~\ref{sec:iterativefit}. 
We have checked for the effect of applying the $2\sigma$ rejection algorithm (as described in 
\S~\ref{subsec:2sig}) on the \bndist s from simulated spectra 
and found that, given that there are a few outliers and no metal contamination, the effect is negligible.
Therefore, we decided not to apply the $2\sigma$ rejection algorithm to simulated \bndist s. 

In Figure~\ref{fig:bnsim} we compare the \bndist s and the respective 
cutoff fits of data (with metal lines excluded, see section 
\ref{sec:metalmask}) and mock spectra at all redshift bins. In both data and simulations, a 
cutoff in the distribution is evident. We also overplot the best fit cutoff (red) and
the 68\% confidence regions (light blue) determined by bootstrapped fits, as described in \S~\ref{sec:data_cutoff}. 
To illustrate the similarities of data and models, the model shown at each redshift is one that has $T_0$ and $\gamma$ 
closest to our final measurement presented in \S~\ref{sec:measurements}.

The main difference is that the \bndist{} of the
data exhibits more lines underneath the cutoff, i.e. 
in the low $b$ and low $N_{\text{HI}}$ part of the panels in Figure~\ref{fig:bnsim}. 
As the SNR distribution is comparable in both diagrams, as well as the amount of
blended absorption systems, we conclude that, if the model assumptions are right, these are most likely metal lines wrongly identified as \lya{} absorption lines. 
Most of these narrow lines are excluded using the 2$\sigma$ rejection described in \S~\ref{subsec:2sig}, 
as indicated by the black dashed lines in the left panels of Figure~\ref{fig:bnsim}. 
This leads to the conclusion that we are able to generate 
\bndist s from our simulations that are similar to those retrieved from data in terms of the cutoff.

\section{Calibration of the Cutoff Measurements}
\label{sec:calibration}
In this section we want to use our simulations to quantify how our cutoff observables
$b_0$ and $\Gamma$ are related to the thermal parameters $T_0$ and $\gamma$. 
Once this calibration is known, it can be applied to our data and, under the assumption that 
simulated and measured \bndist s are similar, 
we can retrieve $T_0$ and $\gamma$ from the data. 

\subsection{Formalism}
To motivate this calibration we start with the 
temperature-density relation \citep{hui1,mcquinn16}, that states that the temperature distribution
as a function of gas density is set by the temperature at mean density
$T_0 = T(\rho_0)$ and the index $\gamma$:
\begin{equation}
 \log T = \log T_0 + (\gamma -1) \log (\rho/\rho_0)\\
 \label{eq:trhorel}
\end{equation}
where $\gamma$ adjusts the contrast level of how much overdensities are
hotter/cooler than underdensities. 

In order to construct a relation between $b_0$ and $T_0$ as well as between 
$\Gamma$ and $\gamma$ we follow the Ansatz presented by \citet{schaye2}.
It states that the overdensity $(\rho/\rho_0)$ and the overdensity in terms of the 
column density $(N_{\text{HI}}/N_{\text{HI,0}})$, where $N_{\text{HI,0}}$ is the column density corresponding to the mean density $\rho_0$, are connected via a power law
 \begin{equation}
   \label{eq:rho_N}
   \log (\rho/\rho_0) = A + B\log (N_{\text{HI}}/N_{\text{HI,0}}). 
 \end{equation}
Furthermore, for absorbers along the cutoff for which turbulent line broadening is negligible,
the line broadening is purely thermal resulting in power law relation between $b_{\rm th}$ and
$T$
 \begin{equation}
   \label{eq:t_b}
   \log T = C + D \log b_{\rm th},  
 \end{equation}
where $b_{\rm th}$ is the thermal Doppler broadening. 
Combining eqns.~(\ref{eq:trhorel}), ~(\ref{eq:rho_N}) and (\ref{eq:t_b}) 
results in a power law relation 
between $b_{\rm th}$ and $N_{\rm HI}$ (eqn.~(\ref{eq:fit_func})) which is the
functional form that we fit to the cutoff of the \bndist{}. 
The coefficients in eqn.~\ref{eq:fit_func} can be written as: 
 \begin{align}
 \label{eq:b0}
  &\log b_0 = \frac{1}{D}(\log T_0 - C + A(\gamma-1))\\
  &(\Gamma - 1) = \frac{B}{D}(\gamma-1).
 \end{align}
Eqn~(\ref{eq:fit_func}) represents the line of minimal broadening at a given column density \NHI{} (therefore $b_{\rm th}$), because absorbers in this relation are
strictly thermally broadened.
If the normalization constant \NHIO{} is chosen so that it represents the column density value of a cloud with mean-density, then 
$A=0$ (see eqn.~(\ref{eq:rho_N})), i.e. the dependency on $\gamma$ disappears from $\log b_0$ in eqn.~(\ref{eq:b0}). 
Taking this into account and 
redefining $\kappa = \frac{D}{B}$ we can re-write these equations as:
\begin{equation}
  \log T_0 = D \log b_0 + C
  \label{eq:calib_relation_slooff}
\end{equation}
\begin{equation}
  (\gamma - 1) = \kappa (\Gamma-1)
  \label{eq:calib_relation_k}
\end{equation}

We can calibrate these relations by fitting the cutoff of mock datasets extracted from our simulations in combination with the same cutoff fitting algorithm 
we applied to the data. 
This approach has the advantage 
that it does not require the assumption that gas is only thermally broadened. Thus we can account for the effects of pressure smoothing 
and thermal broadening on the position of the cutoff in a generalized way.

\begin{figure}
 \plotone{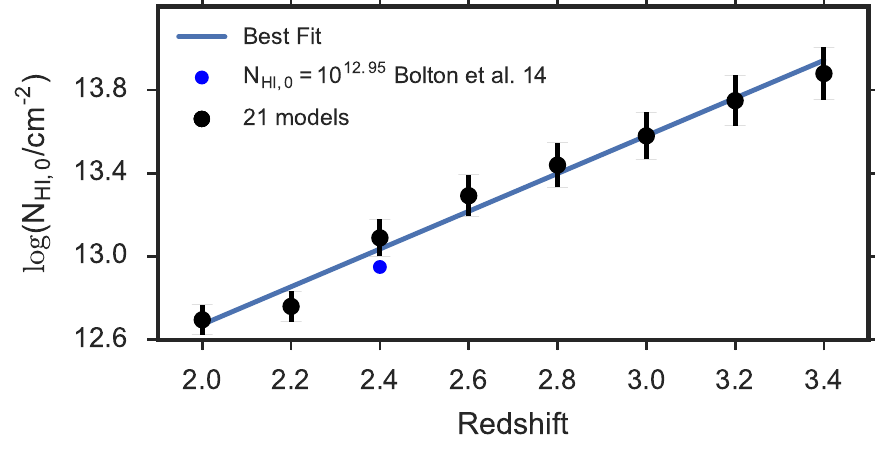}
 \caption{Values of $N_{\text{HI,0}}$(z) from our simulations. 
 The black points are calculated based on the mean flux correction from \citet{becker1} 
 applied to our skewers using eqn~(\ref{eq:boltonNHI}). 
 The error bars reflect the variance in the mean flux re-scaling value (i.e the strength of 
 the UVB) and $T_0$ in the 26 models used in this work. The blue line is a linear fit to the black points, which will be used for estimating $N_{\text{HI,0}}$(z) 
 in this work. For comparison, we show $N_{\text{HI,0}}(z=2.4)$ from \citet{bolton1}, from hydrodynamic simulations.}
 \label{fig:NHI0}
\end{figure}
\subsection{Estimation of $N_{\text{HI,0}}$}
\label{sec:NHI0}
The motivation for normalizing the
\NHI{} values with \NHIO{}, is that it simplifies the calibration
between the $b$-\NHI{} relation and the $T$-$\rho$ relation to be a
one-to-one mapping between $b_0$-$T_0$ and $\gamma-\Gamma$ (equations~(\ref{eq:calib_relation_slooff}) 
and (\ref{eq:calib_relation_k})), with the former
governed by two parameters $(C,D)$ and the latter governed by a
single parameter $\kappa$. In other words, any $\gamma$ dependency is removed from eqn.~(\ref{eq:calib_relation_slooff}).

However, in general the mapping between
\lya{} optical depth and density, and hence between \NHI{} and density
depends on the thermal parameters and the metagalactic photoionization rate $\Gamma_{\text{ion,HI}}$. 
This means that in principle \NHI{} =\NHI{}$(\Gamma_{\text{ion,HI}},T_0,\gamma)$, which can be seen directly from eqn.~(\ref{eq:boltonNHI}), 
as the temperature is a function of $T_0$ and $\gamma$. 
This would require determining \NHIO{} for every single thermal model
in order to calibrate the simple relations of eqns~(\ref{eq:calib_relation_slooff}) and (\ref{eq:calib_relation_k}).
Luckily, eqn.~(\ref{eq:boltonNHI}) illustrates that the thermal 
parameter dependency is quite weak scaling as $T^{-0.22}$. 
Instead, the primary dependency is on $\Gamma_{\text{ion,HI}}$. 
Furthermore, because one always adjusts the mean UVB to
give the same mean flux for different thermal models, the
variation of \NHI{} with thermal parameters is even further
reduced. 

The approach that was used in \citet{rudie1} to compute \NHIO{} 
was to adopt a fixed value of $\Gamma_{\text{ion,HI}}$ and compute 
$N_{\text{HI}}$ analytically, i.e. \NHIO{} = \NHI{}$(\Delta=1)$. \citet{bolton1} instead adopted the average value 
of \NHI{} associated with gas at mean density in his simulations. 
In this work we compute \NHIO{} analytically using eqn~(\ref{eq:boltonNHI}) evaluated at mean-density, 
i.e. $\Delta =1$, for the parameters $\Gamma_{\text{ion,HI}}$ and $T_0$ from our simulations. 
Note that we use the effective UV background $\Gamma_{\text{ion,HI}} = \Gamma_{\text{ion,HI,sim}}/A_r$, 
because our simulations were re-scaled 
to give the correct mean flux at a given redshift (see section \ref{subsec:skewers}). 
Figure~\ref{fig:NHI0} shows the average and 1$\sigma$ range of our \NHIO{} values over all of our
thermal models as a function of redshift. This confirms that the variation of \NHIO{} over the different
thermal models is small, as also argued by \citet{bolton1}.

Finally, we applied a fit to the mean values of $N_{\text{HI,0}}$ over the 26 different simulations 
taking the standard deviation as an estimate for the error. 
The best fit linear function has the form 
\begin{equation}
\log({N_{\text{HI,0}}/ \text{cm}^{-2}}) (z) = a (1+z)+c
\label{eq:NHI0}
\end{equation}
with $a=0.6225$ and $c=11.1068$. 
Throughout this work we will use this function to compute $N_{\text{HI,0}}$ values at fixed 
redshifts. 

Our best fit value of \NHIO{} at $z=2.4$ $N_{\text{HI,0}} \simeq 10^{13.22}$ \cmtwo{} 
is inconsistent with the value measured by \citet{bolton1} $N_{\text{HI,0}} = 10^{12.95}$ \cmtwo{}, presumably because 
of the high values of $\Gamma_{\text{ion,HI}}$ they needed to match the opacity measurements by 
\citet{BeckerandBolton2013}. Part of this possible discrepancy could be due to the lower temperature in 
\citet{bolton1}, but the dependency of \NHI{} on $T_0$ is too small to drive 
this difference. While \citet{bolton1} simulations require a value of $\Gamma_{\text{ion,HI}}/10^{-12}\text{s}^{-1} = 1.86$ 
to match \NHI{} to optical depth weighted density using Schaye's relation (eqn.~\ref{eq:boltonNHI}), we use the re-scaled values 
of our simulations, which are consistent with \citet{BeckerandBolton2013} to directly calculate \NHIO{}. 
This difference of $\sim 0.3$ dex will certainly lead to inconsistent values of $b_0$, but since the calibration process 
is carried out using the same values of \NHIO{} for both the data and simulations, the calibration will cancel out differences 
due to \NHIO{} when dealing with $T_0$ as long as the scatter due to $\gamma$ dependency in eqn.~\ref{eq:calib_relation_slooff} remains 
small compared to our statistical error in $b_0$. We further discuss this in \S~\ref{sec:previouswork} when we compare 
our final measurements to \citet{bolton1}.

\subsection{Calibration Using Simulations}
\begin{figure}[t]
\plotone{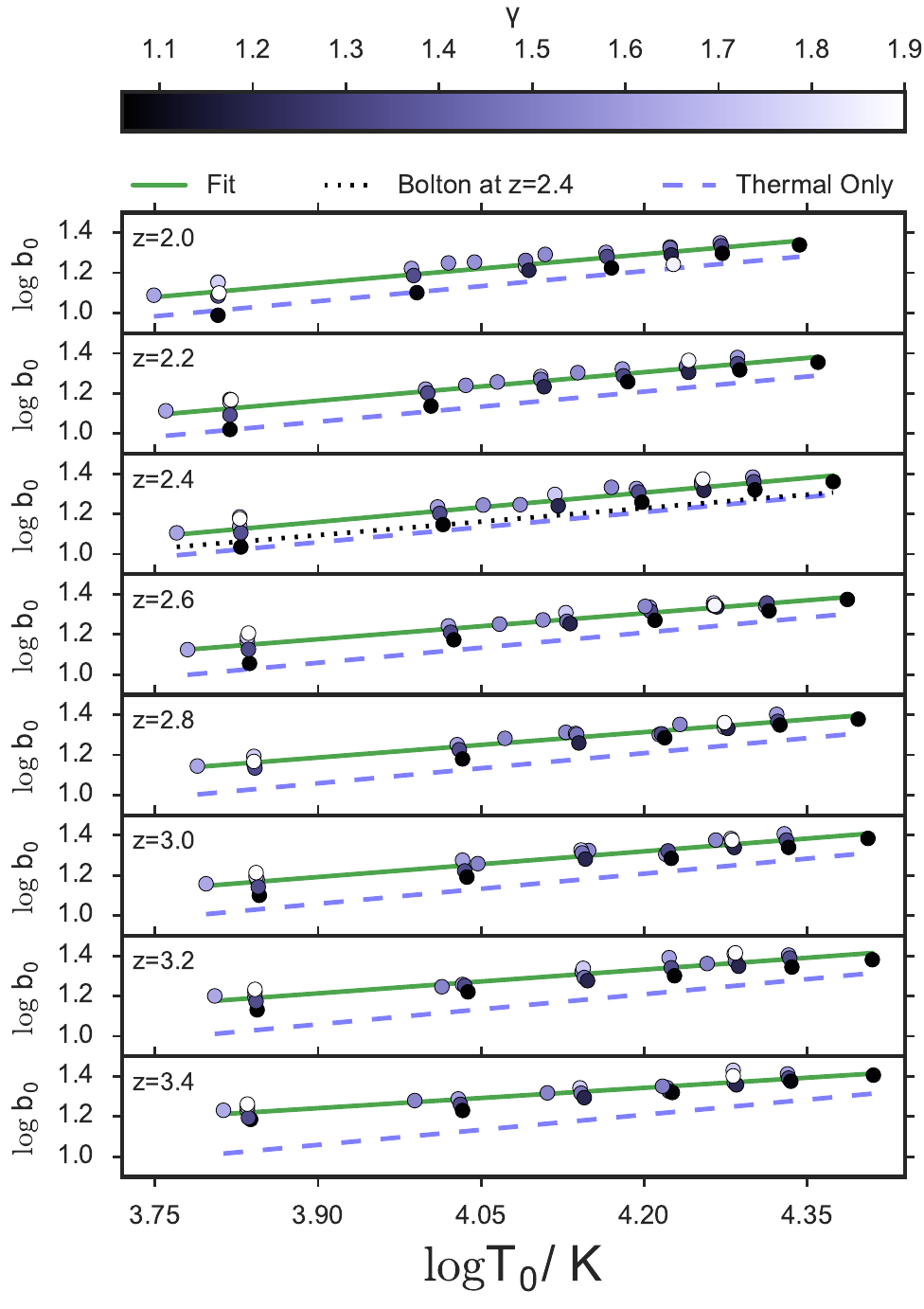}
\caption{Calibration of the $\log b_0$ vs. $\log T_0$ relation. 
 Each point corresponds to a simulated \bndist{}. The points are colored based on their $\gamma$ value. The green line is the best two parameter 
fits to the points. The blue-dashed 
 line represents the case when the value of $b_0$ is due to pure thermal broadening. The scatter is due to unmodelled $\lambda_P$ effects as 
 well as deviations due to $\gamma$-dependency of this relation when $N_{\text{HI,0}}$ does not exactly 
 correspond to the mean-density. At redshift $z=2.4$ we show the line corresponding to the calibration carried out by \citet{bolton1} using hydrodynamic simulations 
 (black dashed).}
  \label{fig:b0_calibration}
\end{figure}

\begin{figure}
\plotone{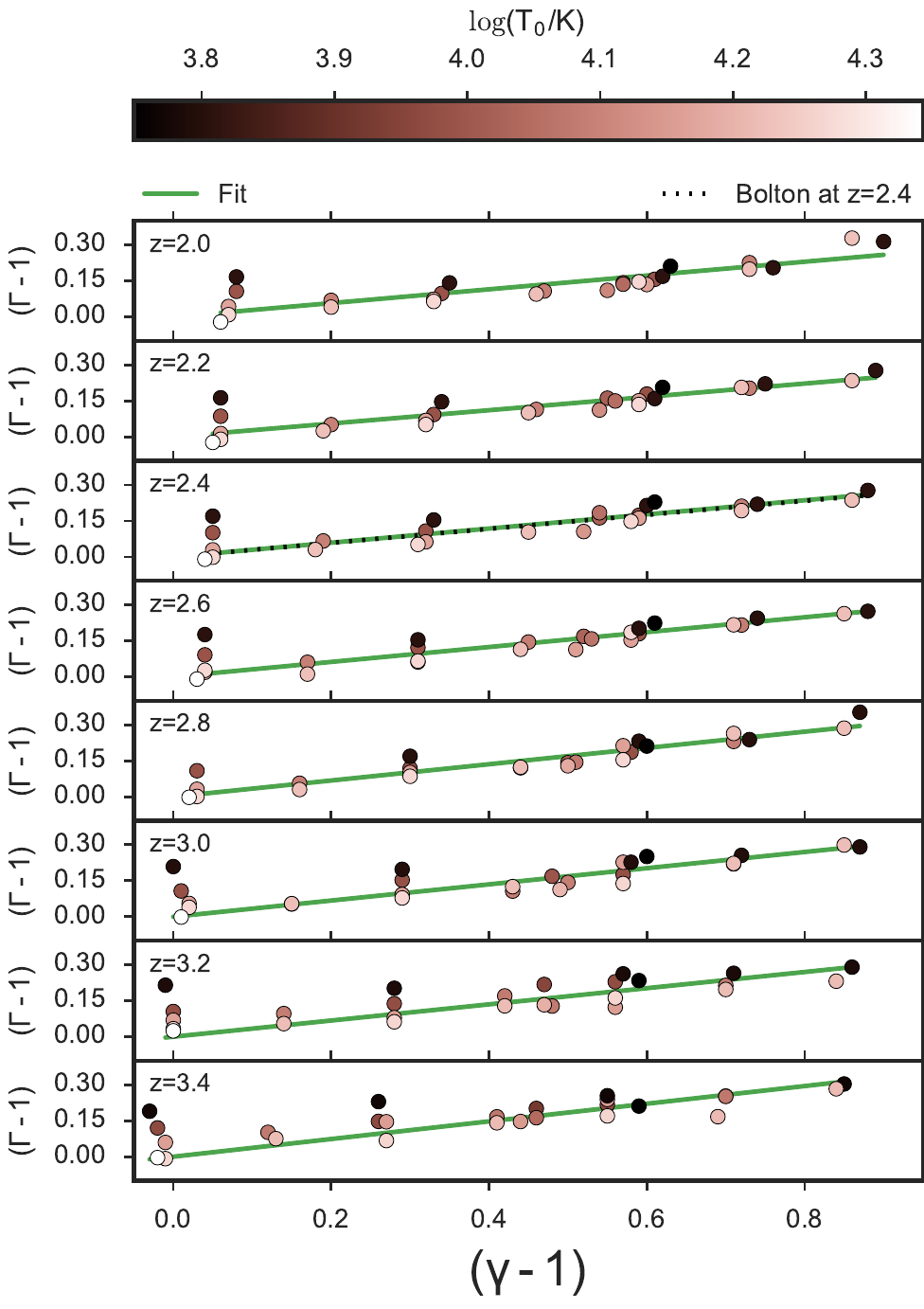}
\caption{Calibration of the $(\Gamma-1)$ vs. $(\gamma-1)$ relation. Each point corresponds to a simulated \bndist{}. 
The points are colored based on their $T_0$ value. The green line represents the best one 
parameter fits to the points. This calibration seem to be independent of the corresponding $T_0$ and $\lambda_P$ values. 
At redshift $z=2.4$ we show the line corresponding to the calibration carried out by \citet{bolton1} using hydrodynamic simulations (black dashed).}
 \label{fig:gamma_calibration}
\end{figure}

In order to generate the calibration between $b_0$-$T_0$ and $\Gamma$-$\gamma$ we ran our cutoff fitting algorithm on simulated \bndist s, each constructed from 100 mock spectra drawn from all of our
26 thermal models at each redshift. The results are shown in Figures~\ref{fig:b0_calibration} and \ref{fig:gamma_calibration}, respectively. 
There we see the simulation input values of $T_0$ and $(\gamma-1)$ for our 26 thermal models plotted against the values of 
$b_0$ and $(\Gamma-1)$ extracted from cutoff fits to each \bndist{}. Each panel corresponds to a different redshift which allows us to capture the evolution 
of the calibration. The green lines are the fits using eqns.~(\ref{eq:calib_relation_slooff}) and (\ref{eq:calib_relation_k}) at every redshift. 
For comparison, we show the calibration of 
\citet{bolton1} at $z=2.4$ in black. In the $\log T_0$-$\log b_0$ diagrams we additionally plot the case in which $b_0$ arises purely due to thermal 
broadening, i.e. $b_0 = \sqrt{2 k_B T_0/m_{\text{HI}}}$. 

The points shown in the diagrams are the median values of $b_0$ and $\Gamma$ from 500 random realizations of the \bndist s with replacement 
rather than the best-fit value of the cutoff parameters of the mock \bndist{}. 
We chose this approach for consistency with how we treated the data, but the results are essentially insensitive to this choice. 

Our $26$ models have different contributions to the thermal broadening $b_0$ due to the different 
values of the pressure smoothing scale $\lambda_P$. 
Similarly, the fact that we assumed one value of $N_{\text{HI},0}$ for all models
with same redshift will introduce a small $\gamma$ dependency in the $\log T_0$-$\log b_0$ relation. 
We want to include our lack of knowledge about $\lambda_P$ and additional effects in the calibration by 
quantifying the amount of scatter that they add into the calibration relations.
This is done by simultaneously fitting equations~(\ref{eq:calib_relation_slooff}) and~(\ref{eq:calib_relation_k}) to the same 2000 bootstrap realizations of 
the points in the $\log T_0$-$\log b_0$ and $(\gamma-1)$-$(\Gamma-1)$ diagrams with replacement. 
The best fit values for every bootstrap realization are stored, giving us the approximated pdfs 
$p(D,C)$ and $p(\kappa)$. 

\begin{figure}
 \plotone{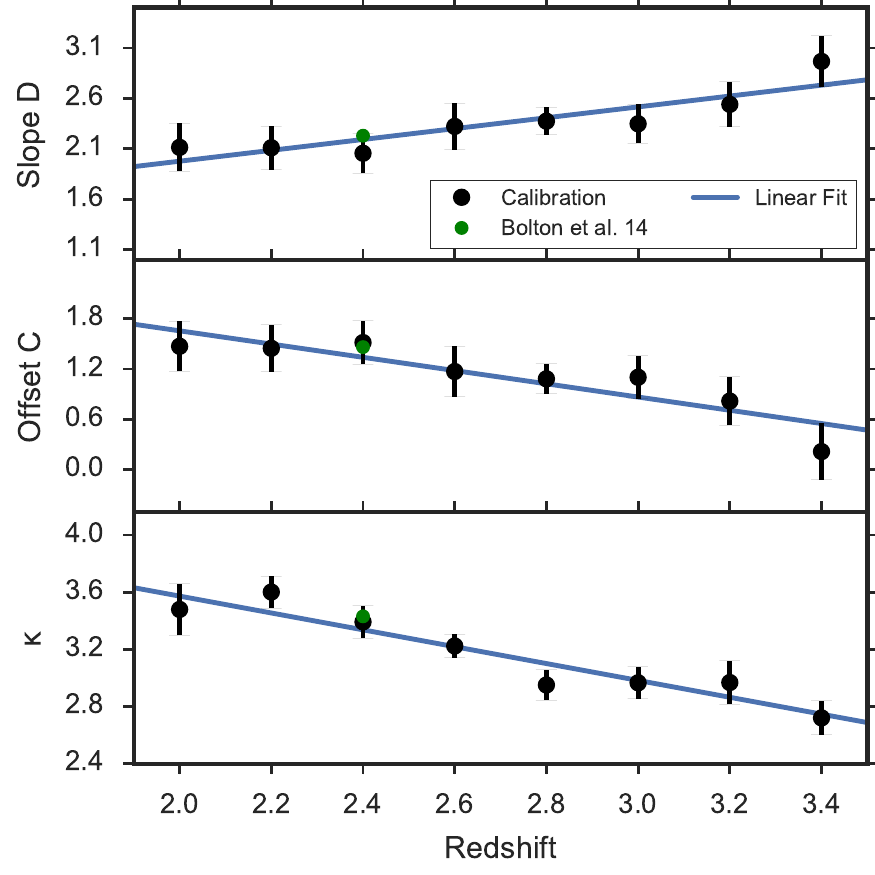}
 \caption{\textbf{Upper panels:} Bootstrapped fit values to the $\log b_0$ vs. $\log T_0$ relation. 
 The error bars reflect the 68\% confidence levels of the marginal distributions of the bootstrapped $p(D,C)$ pdf at each redshift. 
 \textbf{Lower panel} Bootstrapped fit values to the ($\Gamma-1$) vs. ($\gamma -1$) relation. 
 The error bars reflect the 68\% confidence levels of the bootstrapped $p(\kappa)$ pdf at each redshift. The blue lines are linear fits to guide the eye. }
 \label{fig:calibvalues}
\end{figure}

For illustration, the calibration values as a function of redshift are shown in Figure \ref{fig:calibvalues}. 
The error bars correspond to the 68\% confidence intervals of $p(\kappa)$ and the marginal distributions of $p(D,C)$. 
The errors in $\kappa$ are small because the scatter in the $(\gamma - 1)$-$(\Gamma -1)$ relation is 
only slightly driven by dependencies on $T_0$ or $\lambda_P$. 

\begin{figure}[t]
\plotone{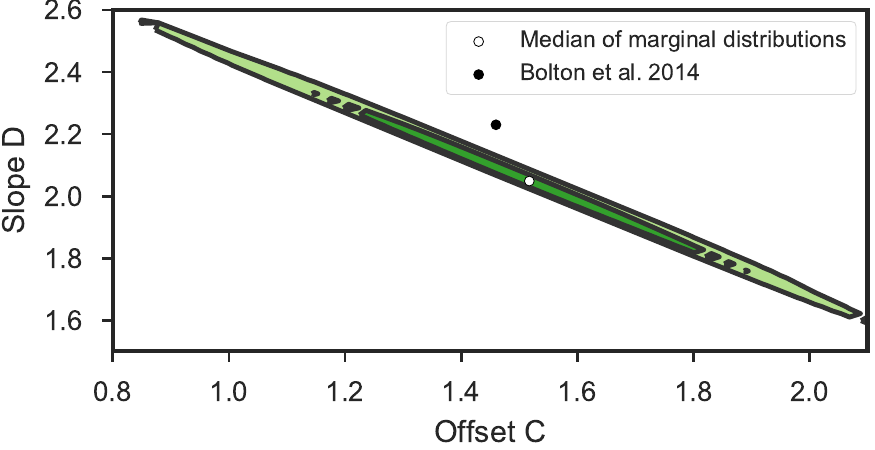}
\caption{Comparison of the $b_0$-$T_0$ calibration values with \citet{bolton1} at $z=2.4$ in terms of the joint distribution of $C,D$. 
The 68\% confidence levels are plotted in dark green and 95\% in light green.}
  \label{fig:jointCD}
\end{figure}

While we agree with the measurements of $C,D$ from \citet{bolton1} at $z=2.4$ in terms of the marginalized distributions 
of $C,D$, his calibration values are about 2$\sigma$ off in terms of the joint PDF $p(C,D)$ as shown in Figure~\ref{fig:jointCD}.
This could be attributed to the difference in method used for cutoff fitting (\citealt{bolton1} uses 
least absolute deviation 
while we use a least-squares minimization approach for the cutoff fitting) as well as the difference in \NHIO{}. 
The calibration constant $\kappa$ between $(\gamma-1)$ and $(\Gamma-1)$ we derived agrees within 1$\sigma$ with the 
value reported by \citet{bolton1}. 

The impact of the calibration differences is further discussed when we compare our $T_0$ and $\gamma$ results 
to previous works in \S~\ref{sec:previouswork}.

\section{Results}
\label{sec:measurements}

\subsection{Evolution of $T_0$ and $\gamma$}
\label{sec:evolution}
Concerning the evolution of $\gamma$, the first conclusion we can draw directly from the data cutoff measurements shown in Figure~\ref{fig:VPmatrix} is
that a positive ($\Gamma-1$) is preferred for all redshift bins. This 
implies, see eqn~(\ref{eq:calib_relation_k}), that a positive temperature-density relation index ($\gamma-1$)
is favored at all redshifts probed. In the $p(b_0, \Gamma)$($z=3$) panel in Figure~\ref{fig:b0Gammapdf} about 4\% of the points in $p(b_0,\Gamma)$ are consistent with $\Gamma<1$. 

\begin{figure*}
  \plotone{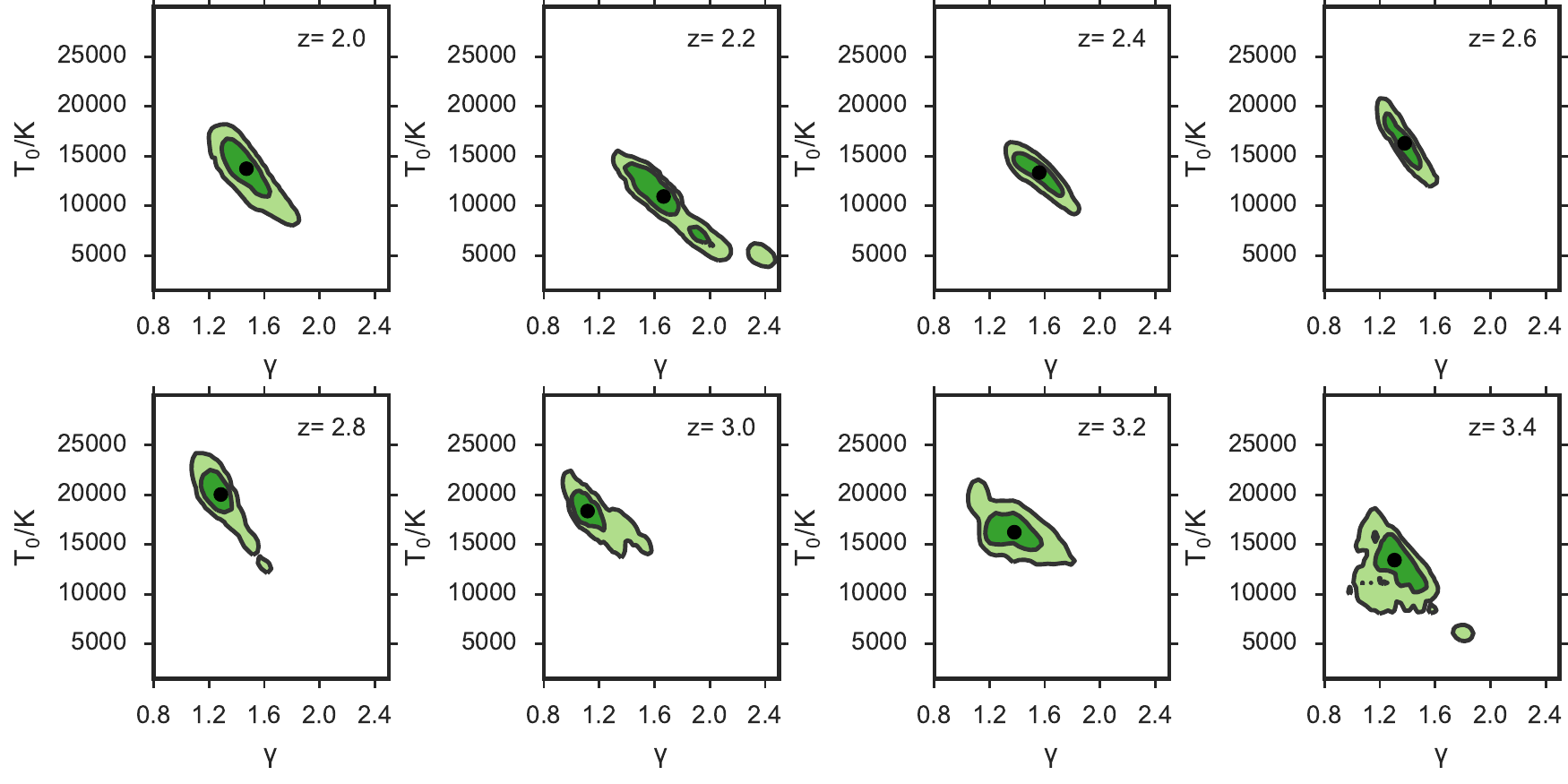}
   \caption{Resulting $p(T_0,\gamma)$ pdfs. This is the combination of our data-measured $p(b_0,\Gamma)$ pdfs with the simulation-extracted 
    calibration $p(\kappa)$ and $p(A,B)$ pdfs. 
    Each panel represents a redshift bin of size $\delta z =0.2$. The 68\% confidence levels are plotted in 
    dark green and 95\% in light green. 
   The black point corresponds to the median of the marginal distributions of $T_0$ and $\gamma$.}
  \label{fig:contours}
\end{figure*}

Having both the cutoff measurements and the calibration in hand, we can now estimate $T_0$ and $\gamma$. It is clear from Figure~\ref{fig:b0Gammapdf} that covariance
in the cutoff fits will lead to a similar covariance between $T_0$ and $\gamma$,
and furthermore, that the scatter in our calibration quantified in Figure~\ref{fig:calibvalues} has to be incorporated into the error budget. To include all of these effects
and arrive at the joint probability distribution $p(T_0,\gamma)$ we adopt
a Monte Carlo approach as follows. 
We combine 
2000 bootstrapped $b_0$ and $\Gamma$ pairs in $p(b_0,\Gamma)$ 
with every single of the 2000 points in the 
bootstrapped calibration pdfs $p(D,C)$ and $p(\kappa)$ from simulations using eqns.~(\ref{eq:calib_relation_slooff}) and~(\ref{eq:calib_relation_k}) 
at every redshift bin. 
The contours 
of the 2000$\times$2000 points in $p(T_0,\gamma)(z)$ estimated via kernel density estimation at every redshift are shown in Figure \ref{fig:contours}.
Comparison with Figure~\ref{fig:b0Gammapdf} indicates that the shape of the $T_0$-$\gamma$ contours are qualitatively similar to the $b_0$-$\Gamma$ contours, 
which results from noise and degeneracy in fitting the cutoff. This is expected from eqns.~\ref{eq:calib_relation_slooff} and \ref{eq:calib_relation_k}. The
contours are slightly broadened by the calibration uncertainty. 
Note that uncertainties in $T_0$ and $\gamma$ are dominated by the statistical errors of $b_0$ and $\Gamma$ due to the high
precision of the calibration process. 

The evolution of the temperature at mean density $T_0$ and index of the temperature-density relation $\gamma$ 
measured in this work is shown in Figures~\ref{fig:thermal_evolution_models} and \ref{fig:thermal_evolution_measure}. 
The error bars are calculated using the 16 and 84 percentiles of the marginal distributions of $T_0$ and $\gamma$ from $p(T_0,\gamma)$. 
The main features are that the temperature at mean-density increases 
from $z=3.4$ to $z=2.8$ (peaking at $T_0 \simeq 20000$ K), while $\gamma$ has its lowest value $\gamma = 1.12$ at $z=3.0$. From $z=2.8$ to $z=2.0$, $T_0$ 
decreases again towards $T_0 \simeq 10000$ K while $\gamma$ increases gradually towards $\gamma \simeq 1.6$.

We tested if the evolution of $T_0$/$\gamma$ is consistent with a peak/dip by comparing $\chi^2$-distributions $P(\chi^2 | {\rm dof})$ of fits to our measurements, where dof is the number of 
degrees of freedom. For this purpose we use a 4-parameter piecewise linear function $f(z)$ of the form 
\begin{equation}
f(z) = 
\begin{cases}
s_1  (z-z_{\rm br}) + o & z<z_{\rm br}\\
s_2  (z-z_{\rm br}) + o & z\geq z_{\rm br},
\end{cases}
\end{equation}
shown in light gray in Figure~\ref{fig:thermal_evolution_models}, that describes two linear function parametrized with 
two slopes $s_1$ and $s_2$, an offset $o$ and a break redshift $z_{\rm br}$. 
For comparison we also compute the best fits for a 2-parameter linear evolution and a constant.
For the evolution of $T_0$, a piecewise 
linear function with a best fit break at $z_{\rm br}=2.9$ results in a $P(\chi^2 | {\rm dof}) = 0.097$ for
4 dof. 
The best fit linear evolution results in $P(\chi^2 | {\rm dof}) = 6.5\times 10^{-4}$ for 6 dof, 
while no evolution in $T_0$ results in 
$P(\chi^2 | {\rm dof}) = 2.4\times 10^{-4}$ for 7 dof. This provides some indication that
our measurements prefer a model with a peak in the temperature. 
In the case of $\gamma$, a piecewise linear function with a break at $z_{\rm br}=3.0$ results 
in a $P(\chi^2 | {\rm dof}) = 0.12$. This is only slightly better than $P(\chi^2 | {\rm dof}) = 0.06$ that we observe for the 
linear evolution model and best fit constant $\gamma = 1.4$ with 
$P(\chi^2 | {\rm dof}) = 0.01$. This suggests that a dip in the evolution of $\gamma$ is slightly preferred given the 
size of our error bars. A comparison of all fits including the reduced $\chi^2$ is given in Table~\ref{tab:fits}.

The peak in $T_0$ is suggestive of a late time $z\sim 3$ process heating the IGM.
The reionization 
of singly ionized helium \ion{He}{2} (\ion{He}{2} $\rightarrow$ \ion{He}{3}) 
by a QSO driven metagalactic ionizing background is the most obvious
candidate that would produce such an effect. 
It has also 
been argued that HeII reionization ends around $z \sim 3$ \citep{worseck1}, which coincides 
with the redshift at which our measurements of $T_0$ appear to peak
\citep{Sanderbeck, puchwein2, jose1}. 

Additionally, if the temperature increase comes about independently of the density of the IGM, 
i.e. the photoionization rate is much higher than the recombination rate everywhere, 
then the IGM is driven to a temperature-density relation that is close to isothermal (see non-equilibrium simulations in \citealt{puchwein2}). 
This causes a flattening of the temperature-density relation, which corresponds to a dip in the evolution of $\gamma$. 
In case that the amount of heating is proportional to the neutral fraction of the gas, e.g. high density regions with higher 
recombination rate experience more heating, then the flattening of $\gamma$ is expected to be less prominent \citep{puchwein2}. 
Given that our data only slightly prefers a dip in $\gamma$ over a constant evolution, 
we can not clearly disentangle these scenarios. Furthermore, the evolution of $\gamma$ seems to be consistent 
with a constant if we apply a least absolute deviation method for the cutoff fitting (see Appendix~\ref{app:alberto}).

\begin{deluxetable}{lcclcc}
\tablecolumns{5}
\tablecaption{Goodness of fit for different models. \label{tab:fits}}
\tablehead{\colhead{Function}& Param. &\colhead{dof}&\colhead{$P(\chi^2 | {\rm dof})$}&\colhead{$\chi^2_{\text{red.}}$}}
\startdata
Constant          & $T_0$    & 7 & $2.4 \times 10^{-4}$ & 3.67 \\
                  & $\gamma$ & 7 & 0.01                 & 2.13 \\\hline
Linear            & $T_0$    & 6 & $6.5 \times 10^{-4}$ & 3.56 \\
                  & $\gamma$ & 6 & 0.06                 & 1.42 \\\hline
Piecewise Linear  & $T_0$    & 4 & 0.097                & 1.30 \\
                  & $\gamma$ & 4 & 0.12                 & 1.11
\enddata
\tablenotetext{a}{Fit for different model types (first column) to the evolution of the parameters 
of the temperature-density relation (second column) 
measured in this work. The goodness of the fit is expressed as the value of the $\chi^2$ distribution given the number of degrees 
of freedom (dof, third column), $P(\chi^2 | {\rm dof})$ (fourth column). Additionally we show the reduced $\chi^2$ (fifth column).}
\end{deluxetable}
After HeII reionization and its concomitant heat injection are
complete, the IGM is expected to cool down on a timescale
of several hundred Myr \citep{hui1, mcquinn16}, or $\Delta z \sim 1.0$, 
and asymptote to a $T_0$ and $\gamma$ set by the interplay of the photoionization heating and 
adiabatic cooling, independent of the details of reionization. 
Due to this process, the IGM is heated by photoionization and then 
left to cool by cosmic expansion once most of the \ion{He}{2} is ionized. This physical picture is consistent 
with our measured evolution of $T_0$ and $\gamma$. 

\subsection{Comparison with Models}

\begin{figure*}
  \plotone{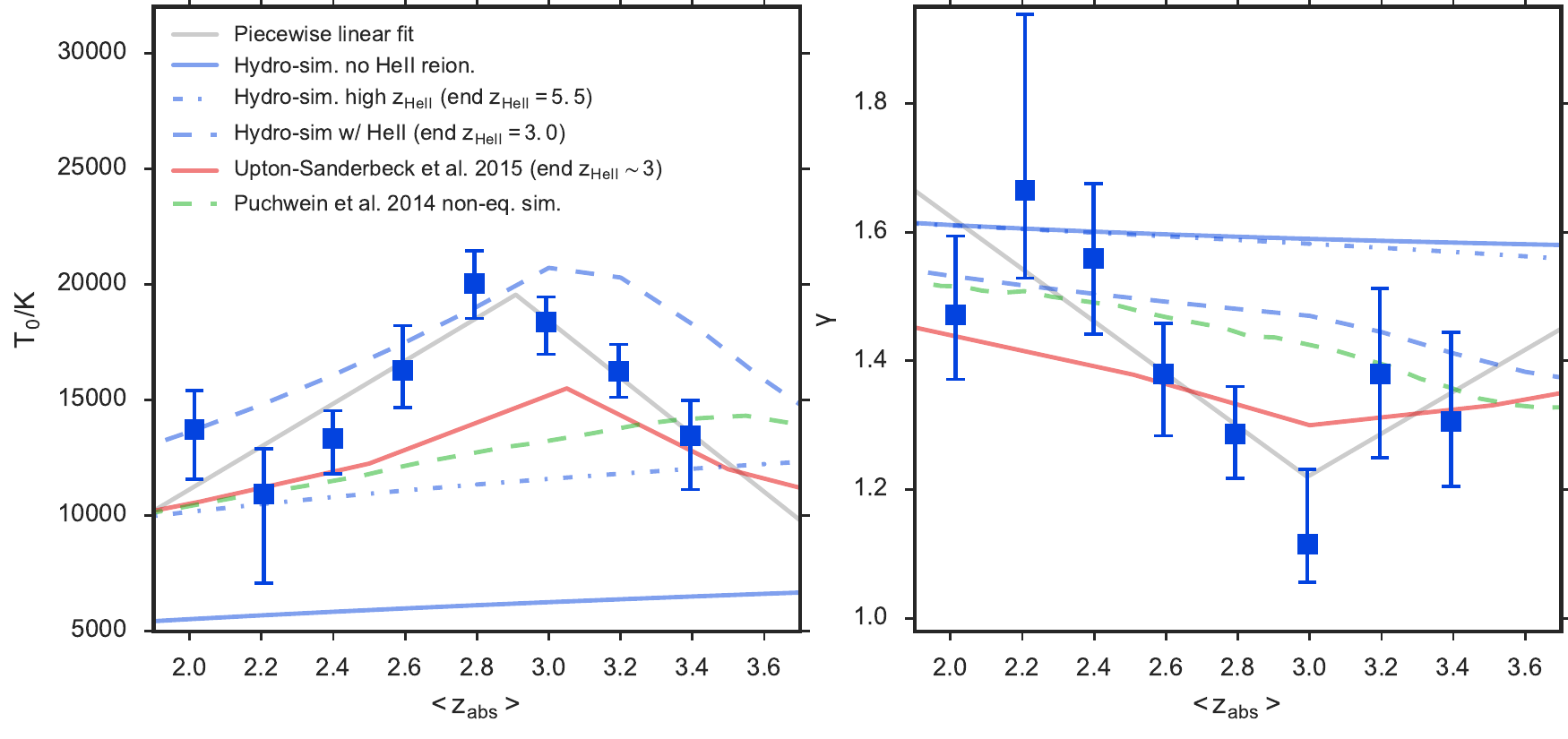}
  \caption{Evolution of $\gamma(z)$ and $T_0(z)$ compared to models. 
  The measurements come from the marginal distributions of $p(T_0, \gamma)$ generated by combining all points in the bootstrapped 
  pdf $p(b_0, \Gamma)$ from the data cutoff fits with all points in the calibration pdfs $p(C,D)$ and $p(\kappa)$ using eqns.~\ref{eq:calib_relation_slooff} 
  and \ref{eq:calib_relation_k}. The error bars are estimated using the 16 and 84 percentiles of the marginal distributions of $p(T_0, \gamma)$. 
  For comparison we plot 3 different Nyx simulations from \citet{jose1}, a semi-analytical models by \citet{Sanderbeck} and a non-equilibrium reionization simulation 
  by \citet{puchwein2}. A best fit 4-parameter piecewise linear function (described in \S~\ref{sec:evolution}) is shown in light gray.}
  \label{fig:thermal_evolution_models}
\end{figure*}
In Figure \ref{fig:thermal_evolution_models}, we compare our measurements to a semi-analytical model 
by \citet{Sanderbeck} constructed by following the photoheating history of primordial gas (red solid line) and 
non-equilibrium reionization simulations by \citet{puchwein2}.
We also compare to different thermal histories from 
the THERMAL suite \citep[blue curves from Nyx simulations,][]{nyx, Lukic2015}. Each Nyx simulation was run 
using different UVB and applying different heat inputs to create three different thermal histories following the method introduced in \citet{jose1}: 
(1) No \ion{He}{2} reionization (blue solid line) 
(2) \ion{He}{2} reionization ending at z =3 with a temperature input 
$\Delta T_{\text{HeII}} = 3 \times 10^4 $K (blue dashed line) and 
(3) \ion{He}{2} reionization ending at $z =5.5$ with a temperature input 
$\Delta T_{\text{HeII}} = 1.5 \times 10^4 $K (blue dot-dashed line).

First we note that if \ion{He}{2} reionization never happened or ended at high redshift, then the simulations suggest 
that $T_0$ would be $\sim 10000$K lower than our measurements at $z = 3$. 
Furthermore, in agreement with the models, 
the temperature at mean density decreases at $z < 3$. Our measurements suggest that $T_0$ is higher than the 
\citet{Sanderbeck} fiducial model and \citet{puchwein2} non-equilibrium simulation, with the difference that 
the non-equilibrium simulation peaks at higher redshift.

The evolution of $\gamma$ from \citet{Sanderbeck} shows a dip at $z = 3$ nearly at the same position as our 
lowest measurement. The $\gamma$ dip in the non-equilibrium simulation appears at higher redshifts, coinciding with the corresponding peak in $T_0$. 
The thermal evolution of the Nyx simulation (2), with \ion{He}{2} reionization at $z=3$, shows a larger $\gamma$ 
at this redshift because the heating due to \ion{He}{2} reionization in the model is more extended and already started at higher 
redshift (see \citealt{jose1} for more details on the models and their intrinsic limitations). In
summary, our measurements of $T_0$ are suggestive of a heating event taking place
between $z=3.4$ and $z=3$.

\subsection{Comparison with Previous Work}
\label{sec:previouswork}

\begin{figure*}
  \plotone{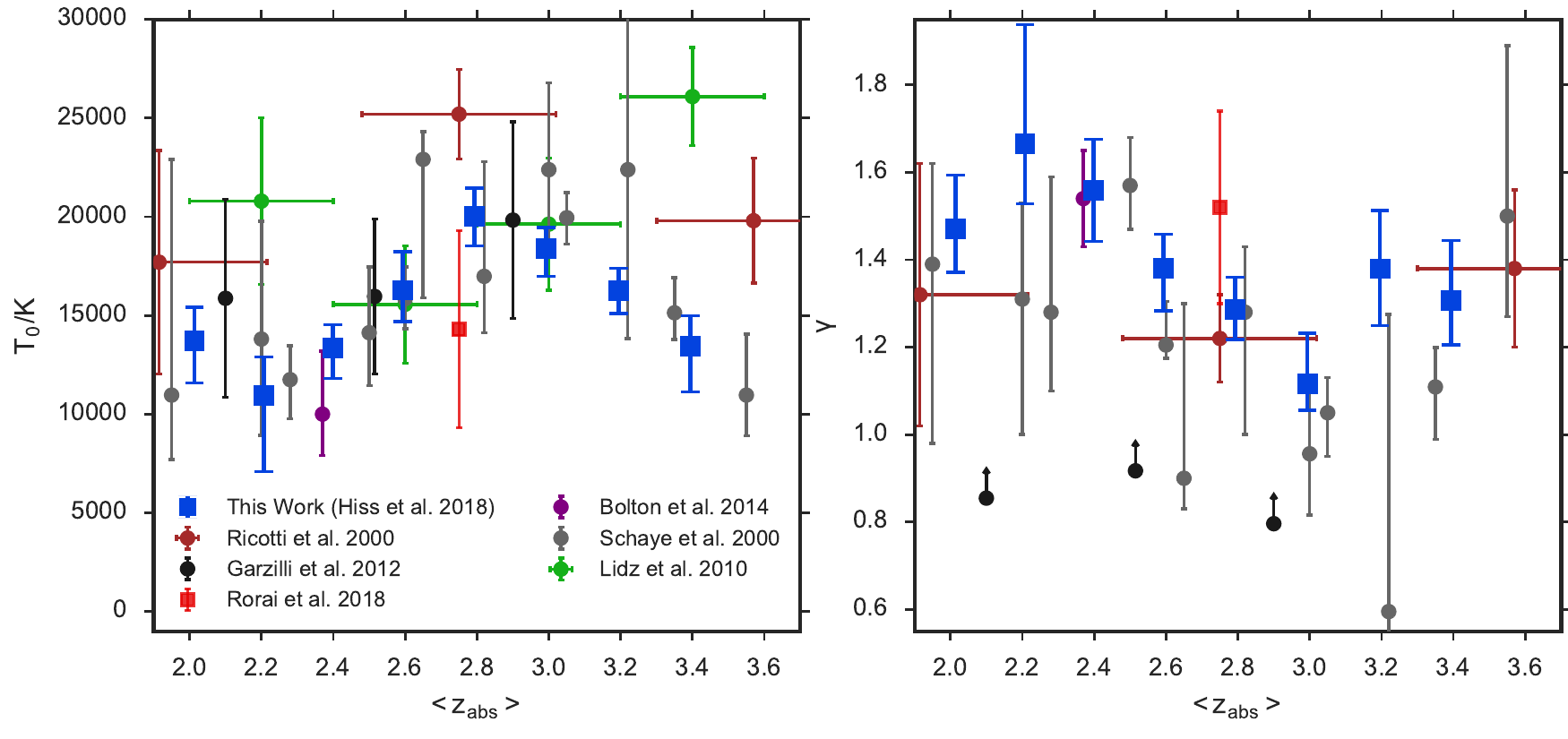}
  \caption{Evolution of $\gamma(z)$ and $T_0(z)$, based on the marginal distributions of the $p(T_0,\gamma)$ pdfs, 
  compared to previous measurements.}
  \label{fig:thermal_evolution_measure}
\end{figure*}
We can directly compare our cutoff fitting results at $z=2.4$ with those presented in \citet{rudie1}, shown 
in the $z=2.4$ panel of Figure~\ref{fig:VPmatrix}.
At $z=2.4$, our bootstrapped cutoff position measurement yields $\Gamma = 1.17 \pm 0.03$, which is in good agreement with 
$\Gamma = 1.156 \pm 0.032$ measured by \citet{rudie1}. If we evaluate their measurement 
$b_{0R} =  b(N_{\text{HI,0}} = 10^{13.6}~$\cmtwo{}$) = 17.56 \pm 0.4$ km/s 
at the position of our $N_{\text{HI,0}}(z=2.4) = 10^{13.22}$~\cmtwo{} while keeping their $\Gamma$ fixed, this measurement 
becomes $b_{0R}' = 15.32 \pm 0.55$ km/s.
Our measurement $p(b_0, \Gamma)$ marginalized over $\Gamma$ (with $b_0 = 18.68 ^{+0.74}_{-1.07}$ km/s) is 
more than $3\sigma$ higher than this value, indicating tension between our measurements and \citet{rudie1} in terms of $b_0$. 
This discrepancy is probably due to a different implementation of the cutoff and VP-fitting 
algorithms used. We performed a cutoff fit our data at $z=2.4$
using a least absolute deviation algorithm and although it tends to lead to smaller values of $b_0$, we 
can not reproduce this low cutoff.

The left panel of Figure \ref{fig:thermal_evolution_measure} shows a comparison of our $T_0$ evolution with previous measurements. 
Our measurements of $T_0$ are in good agreement with those of \citet{schaye1}. We disagree 
with \citet{ricotti1} at $z>2.4$, where we tend to measure significantly lower temperatures.

Note that our $T_0$ measurement agrees with \citet{bolton1}, who recalibrated the cutoff measurement of 
\citet{rudie1} at $z=2.4$. The fact that we measure inconsistent values of $b_0$ should lead to inconsistent 
values in $T_0$. However, given the difference in our calibration values $D,~C$, this inconsistency is alleviated. 
Furthermore, \citet{bolton1} added a systematic error contribution to his statistical uncertainty 
in $T_0$ due to scatter in the $N_{\rm HI}$-overdensity relation in his simulations, 
that lead to a 0.2 dex uncertainty in \NHIO{}. 
When adopting values of \NHIO{} that are 0.2 dex above/below the values determined in \S~\ref{sec:NHI0} self consistently 
in our simulations and data, we observe that the calibration compensates for the choice of \NHIO{}, leading to negligible 
changes in the final results. In other words, choosing a higher value of \NHIO{} will increase the value of $b_0$ almost equally 
in the data and simulations. Note that this is only true as long as the $\gamma$-dependency in eqn.~\ref{eq:calib_relation_slooff} 
remains small. Since our uncertainty in $T_0$ is dominated by the statistical error of $b_0$, we adopt no 
systematic uncertainty term for \NHIO{}.

Our measurements are in good agreement with the wavelet amplitude PDF measurements by \citet{Garzilli2}. 
Comparison with wavelet decomposition measurements by \citet{lidz1} 
in our redshift range shows agreement at intermediate redshifts, but $>2\sigma$ disagreement at $z\sim2.2$ and $3.4$. 
An analogous disagreement has been observed previously in \citealt{becker1} (in the context of curvature measurements), but its source remains unclear. 

We show a comparison of our $\gamma$ values with other measurements in the literature in the right panel of Figure \ref{fig:thermal_evolution_measure}. 
Our measurements of $\gamma$ agree with \citet{schaye1} and \citet{ricotti1}. We also observe a low values of $\gamma$ 
at redshifts around $z=3$.

Our measurement of $\gamma$ at $z\simeq2.4$ agrees with \citet{bolton1}. This was expected given the agreement with \citet{rudie1} 
in terms of $\Gamma$. 

We present a detailed comparison of our measurements of $p(T_0,\gamma)$ at $z=2.8$ with those of \citet{Rorai17vpfit} 
in appendix~\ref{app:alberto}.
\begin{figure*}
  \plotone{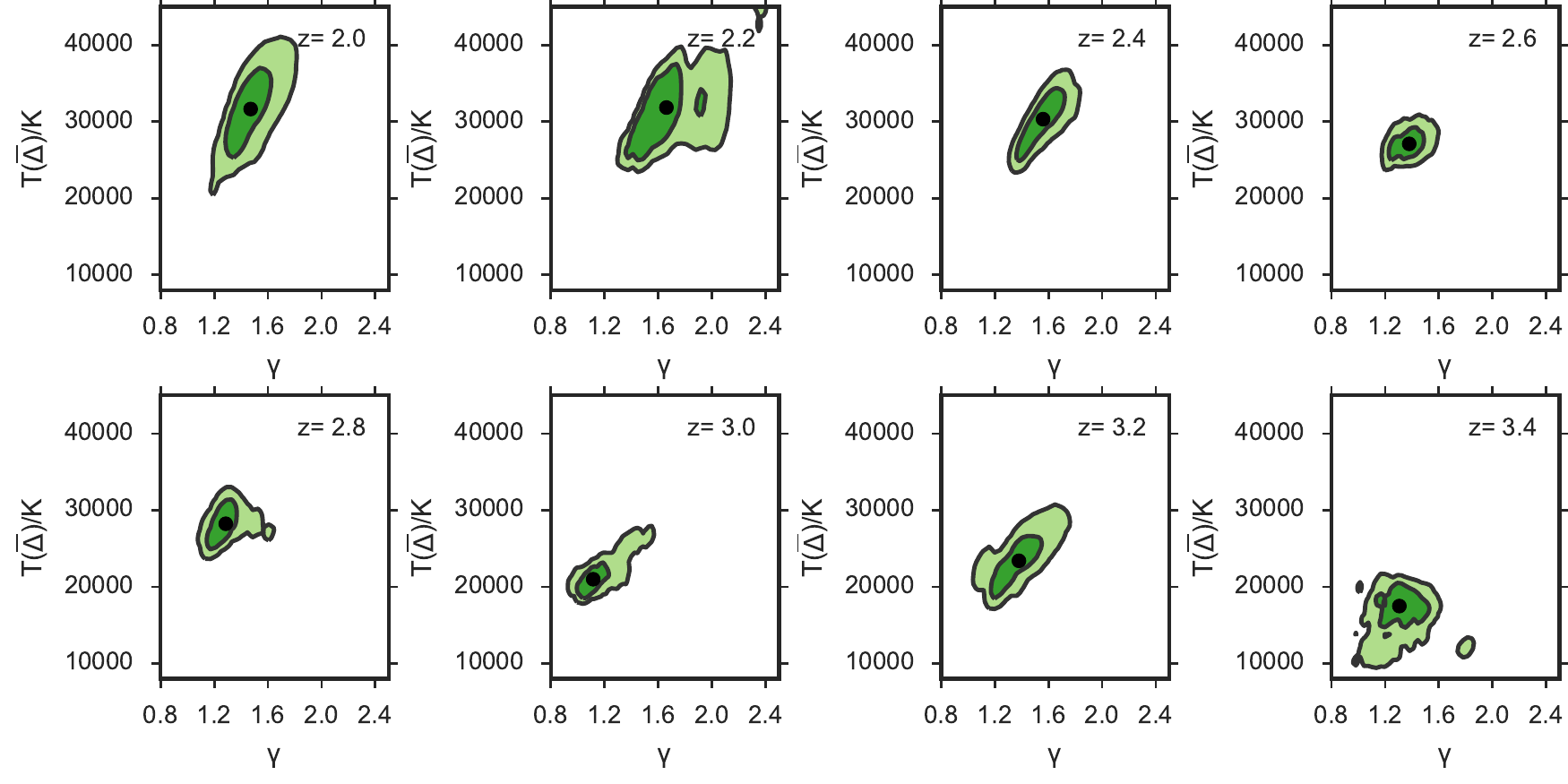}
  \caption{Resulting $p(T(\bar{\Delta}),\gamma)$ pdfs. This is the combination of our calibrated $p(T_0,\gamma)$ pdfs at every redshift with 
  the measurements of $\bar{\Delta}$ by \citet{becker1}. The 68\% confidence levels are plotted in 
  dark green and 95\% in light green. The black point corresponds to the median of the marginal distributions of $T(\bar{\Delta})$ and $\gamma$. }
  \label{fig:contours_becker}
\end{figure*}

\subsection{Evolution of the Temperature at Optimal Density}
The temperature-density relation is traditionally 
normalized at mean-density. 
However, at different redshifts an optical depth of $\sim 1$ in the \lyaf{} traces different overdensities. 
Based on this, \citet{becker1} introduced 
the mean curvature statistic $\left< \left| \varkappa \right| \right>$, 
which is a probe of the thermal state of the IGM that is related to the temperature 
at optimal density $T(\bar{\Delta}) = T(\rho_{\text{opt}}/\rho_0)$ independently of $\gamma$. 

For a fair comparison of our measurements with those from \citet{becker1}, we apply another transformation 
on our measurements so we can look at the evolution of the temperature of the IGM 
in terms of the temperature at the optimal density $T(\bar{\Delta})$.
If we re-write the temperature-density relation in terms of $T(\bar{\Delta})$:
\begin{equation}
 T(\bar{\Delta}) =T_0 \bar{\Delta}^{\gamma-1}
\label{eq:beckerq}
\end{equation} 
we are able to combine our $p(T_0,\gamma)$ pdf with measurements of $\bar{\Delta}$ by \citet{becker1}, 
which have no reported uncertainties. 
Plugging in all pairs of $(T_0, \gamma)$ from $p(T_0, \gamma)$ into eqn.~\ref{eq:beckerq} in combination with a fixed value of 
$\bar{\Delta}$ (linearly interpolated based on \citealt{becker1} to match our redshift bins) allows us to generate 
$p(T(\bar{\Delta}),\gamma)$ pdfs for each redshift. 
This approach takes into 
account any covariance with respect to $\gamma$ in our measurements. 
The resulting $p(T(\bar{\Delta}),\gamma)$ contours are shown in Figure \ref{fig:contours_becker}. 
We note that 
covariance between $T(\bar{\Delta})$ and $\gamma$ is diminished
compared to that between $T_0$ and $\gamma$
(see Figure~\ref{fig:contours} for comparison) 
when taking our measurements to $T(\bar{\Delta})$ space. 
However, note that our $T(\bar{\Delta})$ contours are correlated with $\gamma$ 
in most redshift bins. 

\begin{figure}
\centering
  \plotone{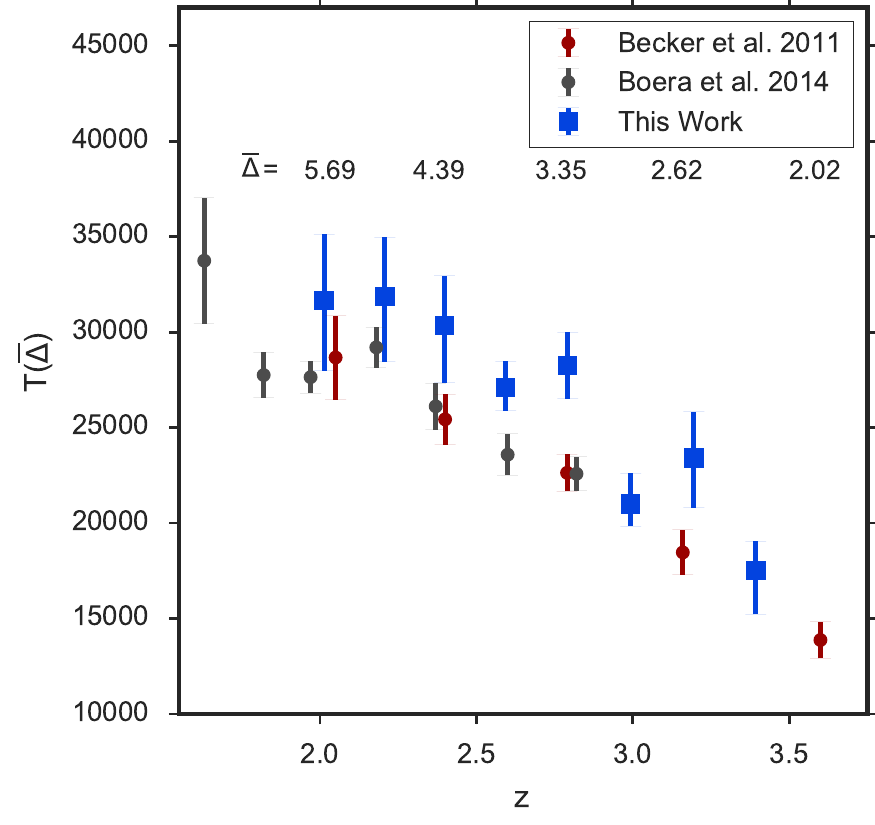}
  \caption{Comparison to \citet{becker1} and \citet{boera1} after combining our $p(T_0,\gamma)$ with \citet{becker1} measurements of $\bar{\Delta}$. We measure a hotter IGM 
  at higher redshifts.}
  \label{fig:becker_comparison}
\end{figure}

Given $p(T(\bar{\Delta}),\gamma)$ joint distributions, we can marginalize out $\gamma$ and compare 
$T(\bar{\Delta}$) directly to \citealt{becker1} and \citealt{boera1} (also computed using the mean curvature method). 
Our 68\% confidence regions for $T(\bar{\Delta})$ as a function of redshift are shown 
in Figure \ref{fig:becker_comparison}.
A comparison with \citet{becker1} is not completely straightforward, given
that the redshift bin sizes are 
different and we are also linearly interpolating their $\bar{\Delta}$ values. 
Broadly speaking, we see agreement with \citet{becker1} and \citet{boera1} at $1\sigma$ level at 
$z<2.4$, $z=3.0$ and $3.4$, 
as well as generally higher temperatures at $2.4\leq z\leq 3.2$ which disagree at the $>2 \sigma$ level. 
Given the method dependency (see Appendix~\ref{app:alberto}) and other systematics associated with cutoff fitting, 
the difference might not be as significant as it appears, once these effects are properly quantified. 
Additionally, if one included uncertainties in $\bar{\Delta}$, it would further alleviate this tension. 
One possible effect that could be playing a role 
is that the curvature statistic is sensitive to metals in the \lyaf{} that do not get
masked, i.e. metal contamination leads to lower values of $T(\bar{\Delta})$ \citep{boera1}.
This effect is potentially more prominent at higher redshifts where blending of \lyaf{} lines
makes it more difficult to identify all metals. Our analysis is in principle less sensitive
to metals given our 2$\sigma$ rejection procedure adopted before cutoff fitting, but the exact
source of this discrepancy remains unclear. 

An overview of all quantities measured and adopted in this work is given in Table~\ref{table:summary}. 
A subset of the measurements on which the distributions $p(b_0, \Gamma)$, $p(D,C)$, $p(\kappa)$ and $p(T_0, \gamma)$ are based is available in machine-readable
form for all redshifts presented and can be obtained in the Zenodo repository \citet{Hiss18_online_material}\footnote{Url: \url{https://zenodo.org/record/1285569}}. 

\begin{deluxetable*}{cccccccccc}
\tablecaption{Measurements and values adopted}
\tablehead{\colhead{$z$} & \colhead{$b_0$} & \colhead{$D$} & \colhead{$C$} & \colhead{$T_0$} & \colhead{$\Gamma$} & \colhead{$\kappa$} & \colhead{$\gamma$} & \colhead{$\bar{\Delta}$} & \colhead{$T(\bar{\Delta})$} \\ \colhead{} & \colhead{(km/s)} & \colhead{} & \colhead{} & \colhead{(K)} & \colhead{} & \colhead{} & \colhead{} & \colhead{} & \colhead{(K)}}
\tablecomments{Summary of all quantities measured/used in this work. The first column shows the center of each redshift bin used. The second column shows the median and percentile based errors of the cut-off fitting parameter $b_0$. The third and fourth columns show the calibration parameters $C,D$ from eqn (\ref{eq:calib_relation_slooff}). The fifth column shows the resulting $T_0$ once the calibration is applied. The sixth column shows the median and percentile based errors of the cut-off fitting parameter $\Gamma$. The seventh column shows the calibration parameter $\kappa$ from eqn (\ref{eq:calib_relation_k}). The eigth column shows the resulting $\gamma$ once the calibration is applied. The ninth column shows the values of the optimal-density that were linearly interpolated from \citet{becker1}. The last column shows the values of the temperature at optimal density $T(\bar{\Delta})$ constructed using eqn. (\ref{eq:beckerq}). \label{table:summary}}
\startdata
2.0 & $18.22^{+0.97}_{-1.39}$ & $2.11\pm{0.245}$ & $1.48\pm{0.305}$ & $13721^{+1694}_{-2152}$ & $1.14^{+0.03}_{-0.03}$ & $3.48\pm{0.185}$ & $1.47^{+0.12}_{-0.1}$ & 5.85 & $31659^{+3690}_{-3455}$ \\
2.2 & $16.89^{+1.37}_{-3.12}$ & $2.1\pm{0.225}$ & $1.46\pm{0.285}$ & $10927^{+1961}_{-3843}$ & $1.19^{+0.07}_{-0.04}$ & $3.6\pm{0.1}$ & $1.67^{+0.27}_{-0.14}$ & 5.1 & $31853^{+3415}_{-3126}$ \\
2.4 & $18.68^{+0.74}_{-1.07}$ & $2.04\pm{0.19}$ & $1.54\pm{0.25}$ & $13334^{+1206}_{-1530}$ & $1.17^{+0.03}_{-0.03}$ & $3.39\pm{0.11}$ & $1.56^{+0.12}_{-0.12}$ & 4.4 & $30335^{+2976}_{-2617}$ \\
2.6 & $20.41^{+1.02}_{-0.87}$ & $2.31\pm{0.23}$ & $1.19\pm{0.3}$ & $16281^{+1940}_{-1601}$ & $1.12^{+0.02}_{-0.03}$ & $3.22\pm{0.08}$ & $1.38^{+0.08}_{-0.1}$ & 3.87 & $27113^{+1234}_{-1372}$ \\
2.8 & $22.67^{+0.55}_{-0.6}$ & $2.38\pm{0.135}$ & $1.08\pm{0.175}$ & $20036^{+1416}_{-1507}$ & $1.1^{+0.02}_{-0.02}$ & $2.95\pm{0.1}$ & $1.29^{+0.07}_{-0.07}$ & 3.35 & $28245^{+1729}_{-1750}$ \\
3.0 & $22.24^{+0.33}_{-0.73}$ & $2.35\pm{0.19}$ & $1.1\pm{0.25}$ & $18371^{+1087}_{-1388}$ & $1.04^{+0.04}_{-0.02}$ & $2.96\pm{0.11}$ & $1.12^{+0.12}_{-0.06}$ & 2.95 & $21002^{+1171}_{-1596}$ \\
3.2 & $21.65^{+0.4}_{-0.52}$ & $2.53\pm{0.23}$ & $0.84\pm{0.3}$ & $16244^{+1153}_{-1135}$ & $1.13^{+0.04}_{-0.04}$ & $2.96\pm{0.145}$ & $1.38^{+0.13}_{-0.13}$ & 2.57 & $23410^{+2623}_{-2409}$ \\
3.4 & $20.8^{+0.71}_{-1.27}$ & $2.97\pm{0.265}$ & $0.22\pm{0.35}$ & $13439^{+1542}_{-2318}$ & $1.11^{+0.05}_{-0.04}$ & $2.72\pm{0.12}$ & $1.31^{+0.14}_{-0.1}$ & 2.3 & $17500^{+2289}_{-1542}$
\enddata
\end{deluxetable*}

\subsection{Caveats}
\label{sec:caveats}
It should be noted that there are a number of assumptions adopted in our work that we summarize as follows.

We assume that the simulated \bndist s are comparable to the ones extracted from the data, or in other words, that the cutoff 
fitting algorithm will respond similarly in both cases. This is a specially problematic assumption, because metals have to be rejected from our 
data which are by construction not present in the simulated mock spectra. Therefore, we observe that the \bndist s from mock spectra generate much 
more concentrated cutoff fitting bootstraps (see Figure~\ref{fig:bnsim}). This effect 
increases the errors measured in $b_0$ and $\Gamma$ in the data, which dominate the error budget of $T_0$ and $\gamma$. 
Furthermore, our simulations do not account for effects such as multimodality 
in the temperature-density relation which could play a role specially at $z>2.8$.

Another assumption is that the calibrations for $T_0$ and $\gamma$ can be done separately, i.e. $p(D,C,\kappa) = p(D,C)p(\kappa)$. 
This is not necessarily true, as these parameters could be correlated. 
As we calculated the calibration values on the same bootstrap samples, any correlation is still preserved. 
We inspected the distributions of $p(\kappa, C)$ and $p(\kappa, D)$ and did not find significant correlation. 

In this work we utilize a least-square fitting algorithm in every iteration of the cutoff fitting process. 
This is a different approach than in previous works and our final results are sensitive to the method chosen. This aspect 
is further discussed in the appendix~\ref{app:alberto} in the context of the comparison of our work with the results of 
\citet{Rorai17vpfit}.

As mentioned in \citet{schaye3}, if the reionization process has large spatial fluctuations and the gas has not settled into one 
temperature-density relation \citep[][see]{compostella1,mcquinn16}, the measurement of the position of the cutoff will be sensitive 
to the gas with the lowest temperature. If this is the case, the temperature measurements should be treated as lower limits to 
the average temperature.
\section{Summary}
\label{sec:discussion}
In this work we assessed the thermal state of the IGM and its
evolution in the redshift range $2.0<z<3.4$ using 75 high SNR and
high resolution Ly$\alpha$ forest spectra from the UVES and HIRES
spectrographs. We exploit the fact that absorbers that are primarily broadened due to the thermal 
state of the gas have the smallest Doppler parameters, which results in a low-$b$ cutoff in the \bndist{}.
We decomposed the Ly$\alpha$ forest of these spectra into a collection of
Voigt profiles, and measured the position of this cutoff as a function of redshift. We calibrate
this procedure using 26 combinations of thermal parameters at each redshift from the THERMAL suite 
of hydrodynamic simulations, 
with different values of the
IGM pressure smoothing scale. We conduct an end-to-end analysis whereby both data and simulations
are treated in a self-consistent way, and uncertainties in both the cutoff fitting, and the calibration procedure
are propagated into our analysis. 

The primary results of this work are: 
\begin{itemize}
\item We see suggestive evidence for a peak in IGM temperature
  evolution at $z\simeq 2.9$. The temperature at
  mean density $T_0$ increases with decreasing redshift over the range
  $2.9<z<3.4$, peaks around $z\simeq 2.9$, and then again decreases
  with redshift over the range $2.0<z<2.9$. 

\item When applying our cutoff fitting procedure, the redshift evolution of $\gamma$ suggests a dip around $z\simeq3.0$ over a linear or constant evolution model 
when using a simple piecewise linear evolution model, that decreases in the redshift interval $2.9<z<3.0$ and 
increases in the interval $2.0<z<3.0$.
        
\item We observe significantly higher temperatures at mean density
  $T_0\simeq 15000-20000\,$K at $2.4<z<3.4$ than the much lower $T_0\simeq6000\,$K
  predicted by models for which \ion{He}{2} reionization did not take
  place, or compared to the $T_0\simeq10000\,$K expected if \ion{He}{2} reionization ended at very high redshift ($z=5.5$).

\item In contrast to previous analyses based on the flux PDF (\citealt{bolton2008, viel1}), 
  our measurements disfavor negative values of $\gamma-1$ at high statistical significance. Assuming that the IGM follows 
  a temperature-density relation closely, this means that inverted temperature-density relations are unlikely at $2.0<z<3.4$. 
  Note that the discrepancies with flux PDF measurements can be attributed to an upturn in temperature at low densities 
  and whether the IGM temperature-density relation is multiphased at low densities \citep{rorai2017}.
 
\item Our measurements of $T_0$ and $\gamma$ can also be phrased as measurements of
  $T(\bar{\Delta})$, which is the quantity measured by curvature studies. We find broad
  agreement with the \citet{becker1} and \citet{boera1},
  curvature measurements at $z < 2.4$, $z=3.0$ and $z=3.4$, but we observe 
  higher values of $T(\bar{\Delta})$ in the interval $2.4\leq z\leq 3.2$.
 
\end{itemize}

In summary, both the suggestive peak in the redshift evolution of $T_0$ at $z\sim 2.9$
and the relatively high IGM temperatures $T\sim 10000-20000$ at $2.0 < z < 3.4$ provide evidence 
for a process that heated the IGM at $z \sim 3-4$. The most likely candidate responsible for this thermal 
signature is \ion{He}{2} reionization. 

In future work we plan to carry out measurements of thermal parameters 
by treating the full probability distribution function of the 
\bndist{}. This method is potentially much more constraining than the 
approach adopted here, which focused exclusively on the cutoff, 
because it utilizes all the information contained in the shape of the 
distribution. Given the existing Hubble Space Telescope Cosmic Origins Spectrograph (HST/COS) ultraviolet spectra \citep[e.g.][]{Danforth2013, Danforth2016} probing 
the $z\lesssim 0.5$ \lyaf{}, this new method could be an interesting 
tool for studying the IGM at lower redshift, especially in light of 
recently reported discrepancies between data and hydrodynamical 
simulations for the distribution of Doppler parameters and column densities 
\citep{Viel2017, Gaikwad2017, Nasir2017}. 

\section*{Acknowledgments}
We thank the members of the ENIGMA group at the MPIA as well as the members of office 217 for the fruitful discussions and 
helpful comments. 
We thank the anonymous referee for carefully reading our manuscript and for their 
insightful comments and suggestions that significantly improved the quality of this work. 
Special thanks to Martin White for providing us with collisionless simulations. 

Some data presented in this work were obtained from the Keck Observatory Database of Ionized Absorbers toward QSOs (KODIAQ), 
which was funded through NASA ADAP grants NNX10AE84G and NNX16AF52G along with NSF award number 1516777.
This research used resources of the National Energy Research Scientific Computing Center (NERSC), which is supported by the Office of Science of the U.S. 
Department of Energy under Contract no. DE-AC02-05CH11231.

Calculations presented in
this paper used the hydra and draco clusters of the Max Planck
Computing and Data Facility (MPCDF, formerly known
as RZG). MPCDF is a competence center of the Max
Planck Society located in Garching (Germany).
\bibliographystyle{apj}
\bibliography{references.bib}
\appendix
\section{QSO Continuum Placement}
\label{appendix}

It is important to note that the continua of the QSOs in our sample
are placed based on the portions of the spectra that have no
apparent absorption and is therefore subject to uncertainty. 
A misplacement of the continuum could
certainly have an
effect on the corresponding optical depth of a line (and therefore on the
line-profile parameters). Different studies show that for high SNR and 
resolution data, the statistical uncertainty of the continuum 
placement is of the order of a few percent at $z<4$ \citep{Kirkman05,Kim07, aldo1,Faucher-Giguere2008b}. 
We assume that our typical continuum uncertainty is of the order $\sim 2\%$/$\sim 5\%$ for 
$z<3$/$z>3$ sightlines. 

To address the effect of continuum misplacement in our study, we
analytically estimate how a 
shift of $2\%$ and $5\%$ in the continuum affects the typical line in our sample. 
This is done by calculating the optical depth at line-center \citep{meiksin1} 
\begin{equation}
\tau_{lc, \text{HI}} \simeq 0.38 \left(\frac{N_{\text{HI}}}{10^{13} \ \text{\cmtwo}}\right) \left(\frac{20 \ \text{km/s}}{b} \right)
\end{equation}
for lines with different column densities and a typical width of 
$b = 19$ km/s, and converting it to flux 
at line-center $F_{lc} = \exp{(-\tau_{lc})}$.
This flux is shifted by $2\%$ and $5\%$ 
to mimic the effect of misplacement of the continuum and then, doing the reverse 
operations and keeping $b$ fixed, we compute the corresponding $\log N_{\text{HI}}$ values. 

For a continuum shift of $2\%$, the corresponding shift in $\log N_{\text{HI}}$ is generally smaller 
than the uncertainty in $\log N_{\text{HI}}$ reported by \vpfit{} within our cutoff fitting range. For a continuum misplacement of $5\%$, 
the \vpfit{} uncertainty becomes comparable to the continuum misplacement effect at column densities 
$\log (N_{\text{HI}}/\text{\cmtwo}) =13$ 
and exceeds it at lower $N_{\text{HI}}$. 

Given the small effects on the column densities at lower-redshift, the errors due to 
continuum placement can be neglected. At lower continuum placement precision ($z>3$) this effect can 
influence the lower column-densities, but our cutoff fitting algorithm is likely not very sensitive to this, 
because it is driven by absorbers with better constrained parameters at higher column densities.

\begin{figure}
\plotone{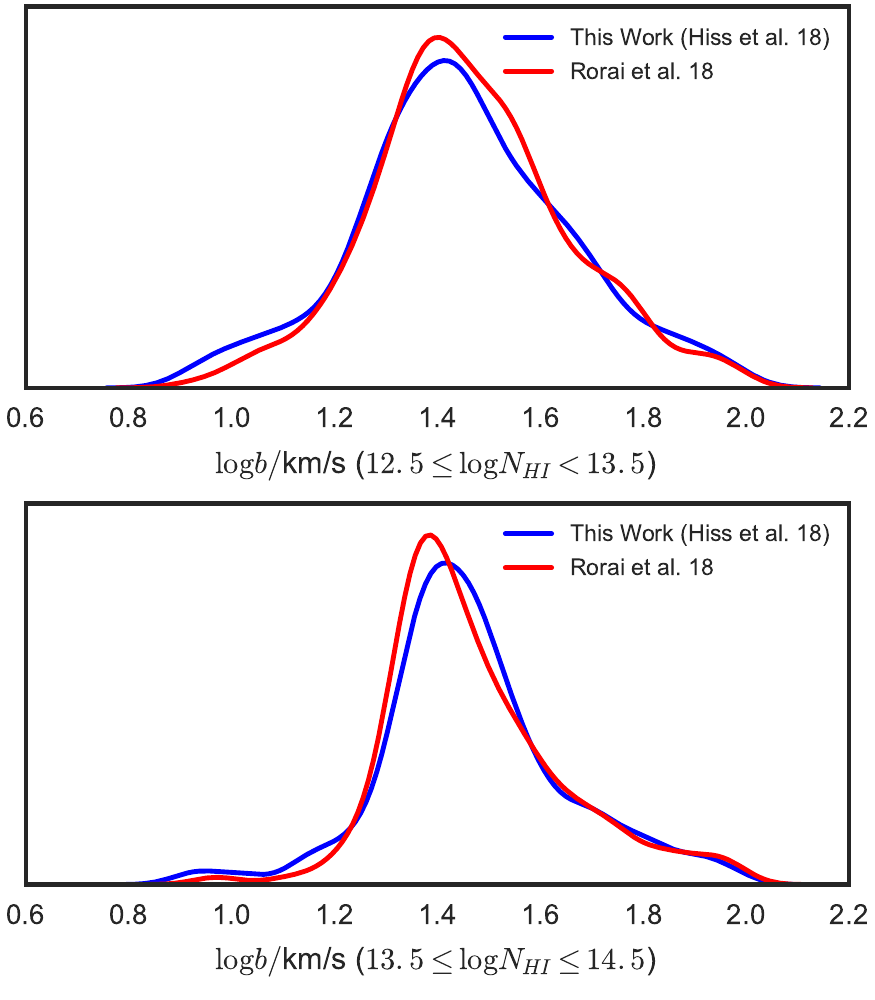}
\caption{Comparison of normalized $\log b$-distributions of data in \citet{Rorai17vpfit} in the 
redshift bin $2.55\leq z\leq 2.95$ and this work in the redshift bin $2.7 \leq z < 2.9$.}
\label{fig:bdist}
\end{figure}

\section{Comparison with R\lowercase{orai et al. (2018)}}
\label{app:alberto}
\begin{figure}
\centering
\plotone{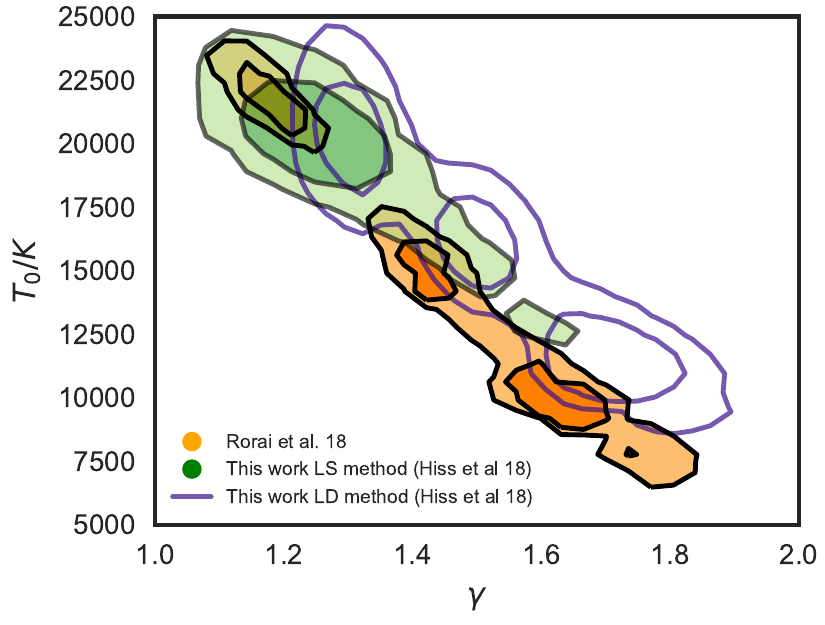}
\caption{Comparison of $T_0$ and $\gamma$ contours in \citet{Rorai17vpfit} and this work 
(LS method in green, LD method in purple) at $z=2.8$. The contours correspond to the 68\% and 95\% confidence regions.}
\label{fig:LScontours}
\end{figure}
Recently, a study by \citet{Rorai17vpfit} reported measurements of the thermal state of the IGM in the redshift interval 
$2.55\leq z\leq 2.95$ which resulted in values of $T_0$ and $\gamma$ that are only marginally consistent with our 
measurement at $2.7 \leq z < 2.9$. To test if the source of this discrepancy originates from the way in which the Voigt-profile 
algorithm was applied to the respective datasets, we plotted the line width distributions for both our line lists 
for two intervals of 1 dex in \NHI{} within the cutoff fitting range. The distributions shown 
in Figure~\ref{fig:bdist} are essentially identical. Thus, any difference 
in the resulting thermal parameters must originate in the cutoff fitting procedure due to contamination, spurious lines 
or differences in the calibration.

In Figure~\ref{fig:LScontours} a direct comparison of the contours of $T_0$ and $\gamma$ shows that 
\citet{Rorai17vpfit} measures a multimodal joint distribution $p(T_0,\gamma)$ (orange) while our measurement 
(green) recovers only the peak with the highest $T_0$ and lowest $\gamma$. 
The main difference between the two methods is that we perform a Least-square (LS) minimization fit at each iteration of the 
cutoff fitting procedure, while \citet{Rorai17vpfit} 
performs a least absolute deviation (LD) fit. 
Our algorithm tends to converge to the peak corresponding to high 
$b_0$ and low $\Gamma$, resulting in this difference. 

\begin{figure}
\plotone{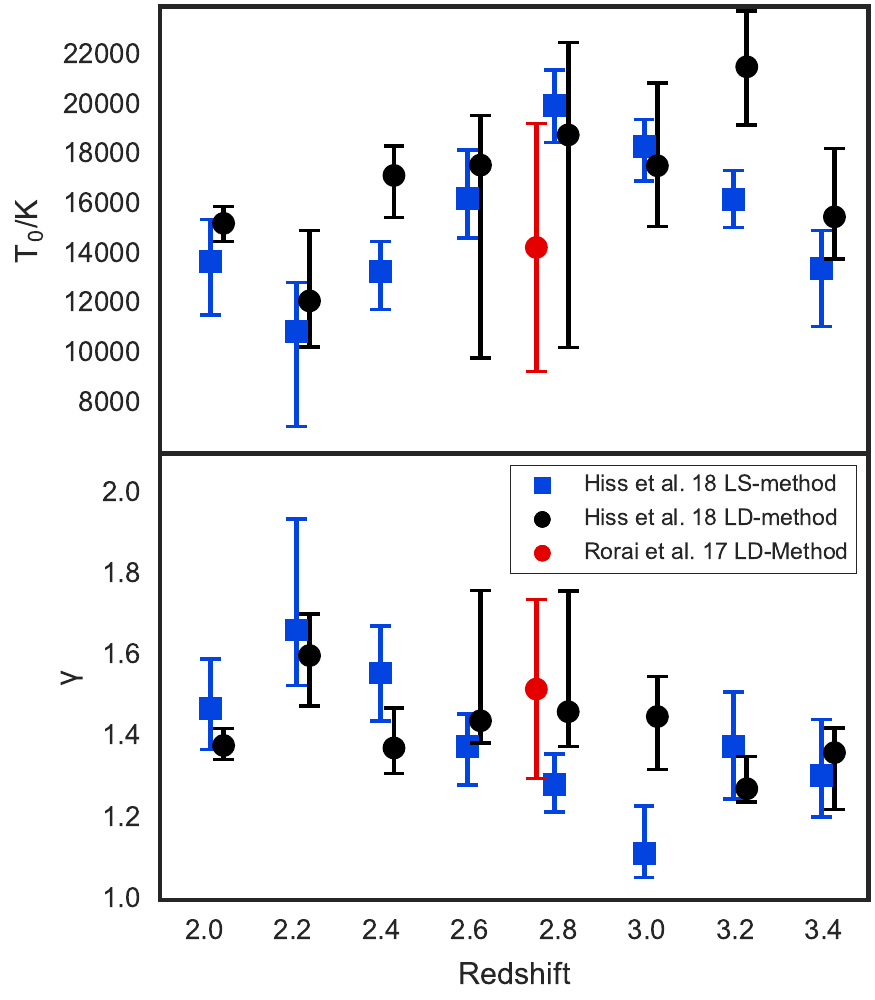}
\caption{Comparison of the marginalized $T_0$ and $\gamma$ in \citet{Rorai17vpfit} (red) and this work (blue). 
We also ran our procedure using a least absolute deviation (LD) minimization cutoff fitting procedure (black). 
The main difference between the methods is that the least-squares (LS) minimization 
method used in this work does not show a multimodal structure at $z=2.6$ and $z=2.8$. Also the evolution of $\gamma$ 
is consistent with a constant, not showing a dip at $z=3$.}
\label{fig:comparisonmethods}
\end{figure}

For comparison we re-run our measurements, this time applying a least absolute deviation 
fit for both our data and simulations. Due to unstable behavior of the least absolute deviation 
method at some redshifts, we applied no 2$\sigma$ outlier rejection (\S~\ref{subsec:2sig}) to our data \bndist{} when applying this method.
We show the resulting $p(T_0,\gamma)$ contours at $z=2.8$ in purple in Figure~\ref{fig:LScontours}. 
The results of the evolution of $T_0$ and $\gamma$ are shown in Figure~\ref{fig:comparisonmethods}. 
Essentially, the main difference between the two methods when applied to our data, is that we see extended uncertainties 
at $z=2.6$ and $z=2.8$, which originate from multimodal distributions $p(T_0,\gamma)$. Furthermore, 
the redshift evolution of $\gamma$ is consistent with a constant $\gamma \simeq 1.4$. 

As in \citet{Rorai17vpfit}, when using the least absolute deviation method, we observe a multimodal 
$p(T_0, \gamma)$ distribution at $z=2.8$ (also at $z=2.6$) in the data that result from a multimodal $p(b_0, \Gamma)$ 
measurement. When dealing with simulated \bndist s both methods lead to unimodal solutions. This opens up the question 
if these multiple peaks in the inference of the cutoff parameters are a real feature due to multimodality in the temperature 
or an artifact of the cutoff fitting procedure due to unknown systematics in the data. Investigating the source of these 
structures is beyond the scope of this paper but we plan to study this in detail in the future.

\begin{figure}
  \plotone{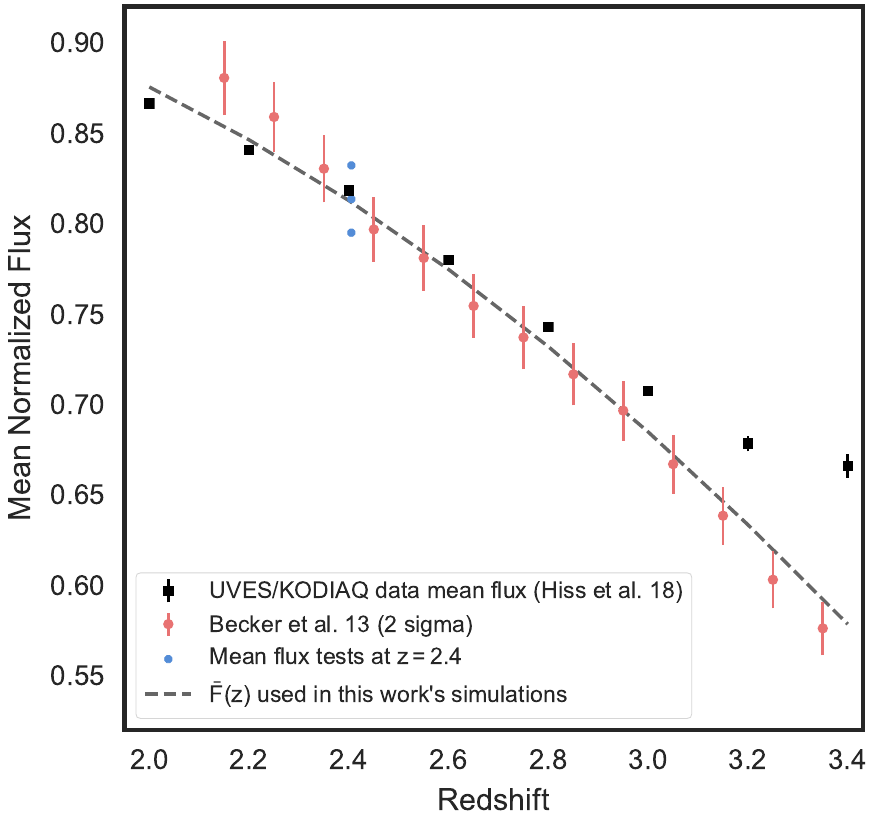}
\caption{Comparison of the mean flux evolution from \citet{jose1} (dashed line, used as a basis for re-scaling the mock skewers 
in this work), the measurements by \citet{Becker_A} (red points) and the mean flux of the data used in this work at each redshift bin. 
The mean flux values used in for the test in Figure~\ref{fig:becker2sig} are shown in blue.}
\label{fig:mean_flux}
\end{figure}
\section{Impact of uncertainties in the mean flux}
\label{app:mean_flux}
We describe in \S~\ref{sec:simulations} how our simulations are re-scaled in terms of flux in order to match the mean flux 
evolution fit $\bar{F}(z)$ from \citet{jose1}. This re-scaling is a standard procedure for
accounting for our lack of knowledge of the precise value of the metagalactic ionizing background 
photoionization rate. 

In Figure~\ref{fig:mean_flux} we show a comparison of the mean flux values inferred from our data set (black squares), 
the values in \citet{Becker_A} (red) 
and the fit to diverse mean Flux measurements from \citet{jose1} (dashed line) which was used as a basis for rescaling the 
mean flux of simulated spectra in this work. 
Only pixels that were not flagged as metals, high column density absorbers or bad 
pixels were used for the calculation of the mean flux in our data. When looking at the mean flux of data, 
we observe that they scatter around the mean flux used in the simulations in the range 
$z = 2.0$ to 3.0. 
\begin{figure}
  \plotone{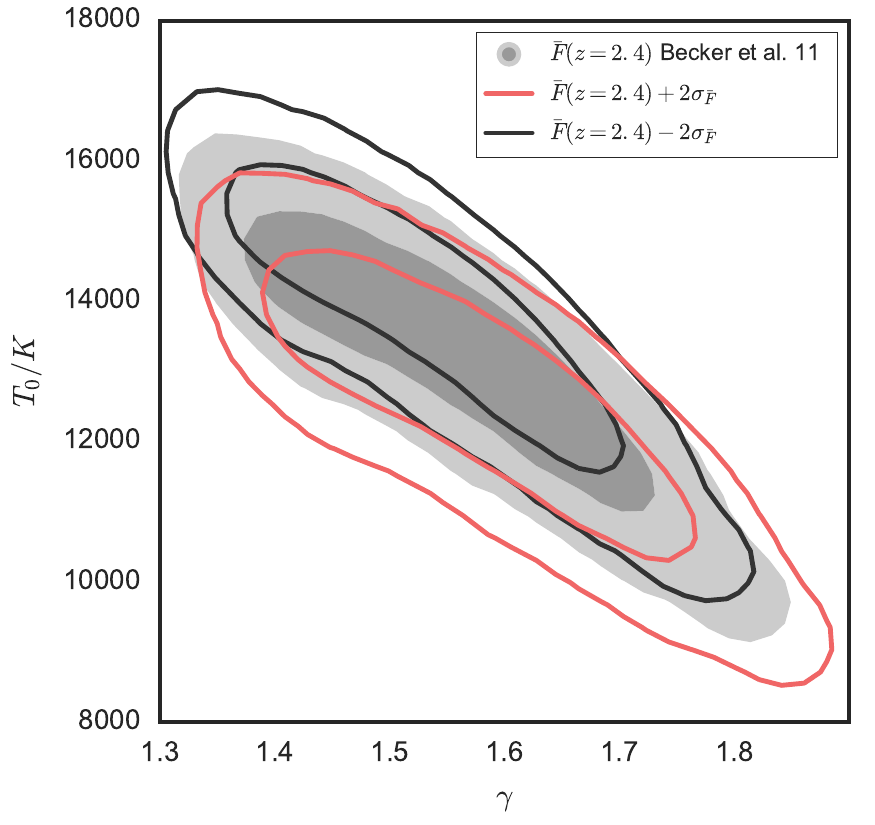}
\caption{Comparison of results at $z=2.4$ for our data calibrated using 
simulations that are scaled to three different mean flux values: 
\citet{becker1} mean Flux (gray filled contours), \citet{becker1} mean Flux +2$\sigma$ (red contour lines) and \citet{becker1} mean 
Flux -2$\sigma$ (black contour lines). The contours correspond to the 68\% and 95\% confidence regions.}
\label{fig:becker2sig}
\end{figure}

To motivate the fact that we do not take into account uncertainties in the mean flux re-scaling 
of our simulations at $2<z<3$, we ran our measurements at $z=2.4$ for different values of the 
flux re-scaling: $\bar{F}$, i.e. our measurement, and $\bar{F} \pm 2 \sigma_{\bar{F}}$, where $\bar{F}=0.8136$ 
is the value interpolated between the measurements of $\bar{F}$ by \citet{Becker_A} at $z=2.35$ and $z=2.45$. For the purpose of being 
conservative, the value of $\sigma_{\bar{F}}$ adopted 
is the error reported by \citet{Becker_A} at $z=2.35$, $\sigma_{\bar{F}}=0.0093$. 
These values are plotted as blue dots in Figure~\ref{fig:mean_flux}.
The corresponding $p(T_0, \gamma)$ measurements are shown in 
Figure~\ref{fig:becker2sig}. Shifting $\bar{F}$ by $2\sigma$ results in a negligible shift of our final results 
at this redshift. 

\begin{figure}
\plotone{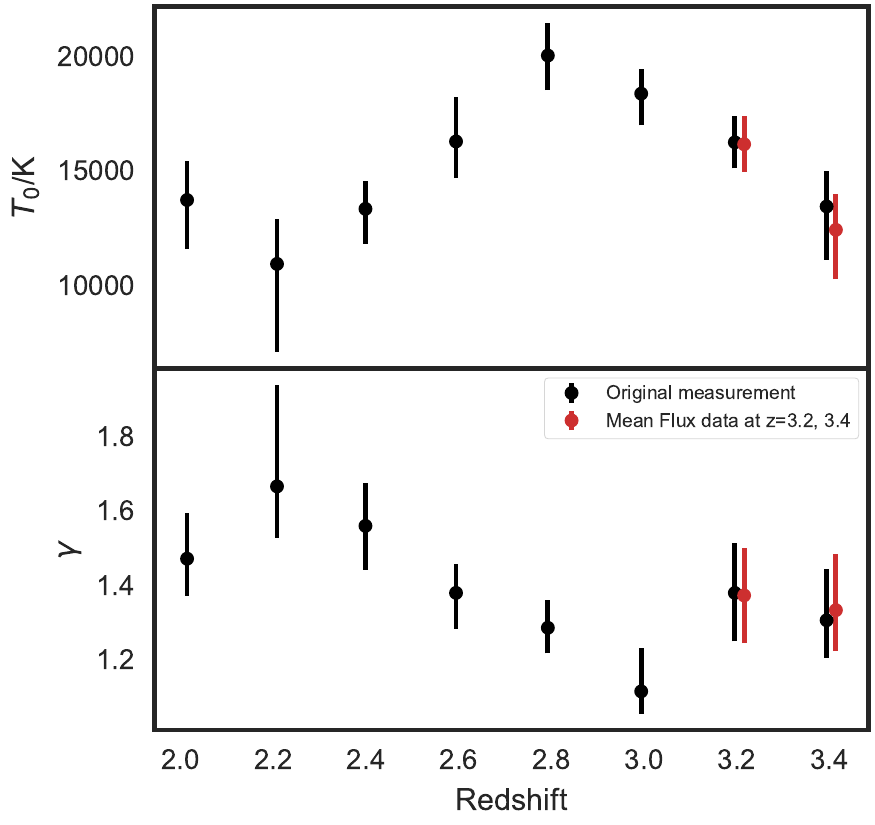}
\caption{Final marginalized $T_0$ and $\gamma$ measurements after re-scaling our models to match the mean flux of our 
data at $z=3.2$ and $z=3.4$ (red) compared to our original measurements (black).}
\label{fig:meanfluxdata2}
\end{figure}

At our highest redshift bins, $z=3.2$ and $z=3.4$ we observe a stronger discrepancy between the mean flux of our models and data.
In order to directly examine the effect of this discrepancy on our measurements, we generated the models used in the calibration 
once again, with the difference that we re-scaled the optical depths to match the mean flux values measured in the data at 
these redshifts. We then applied the calibration based on these new models 
to our cutoff fit results. The results are shown in Figure~\ref{fig:meanfluxdata2}. We observe that the calibrations at these redshifts 
are only slightly sensitive to this change, as our results basically do not change.
\clearpage

\end{document}